\documentclass[%
reprint,
superscriptaddress,
amsmath,amssymb,
aps,
prb,
showkeys,
]{revtex4-2}
\usepackage[T1]{fontenc}
\usepackage{color}
\usepackage{soul}
\usepackage{graphicx}
\usepackage{dcolumn}
\usepackage{bm}
\usepackage{physics}
\usepackage{amssymb}
\usepackage[sort&compress]{natbib}
\usepackage{hyperref}

\usepackage{color}

\usepackage{soul}
\usepackage{todonotes}
\usepackage[normalem]{ulem}

\begin{document}

\preprint{APS/123-QED}

\title{Phase crossover induced by dynamical many-body localization in periodically driven long-range spin systems}

\author{Mahbub Rahaman}
\affiliation{Department of Physics, The University of Burdwan, Golapbag, Bardhaman - 713 104, India}
\author{Takashi Mori}
\affiliation{RIKEN CEMS, 2-1 Hirosawa, Wako, Saitama, 351-0198, Japan}
\author{Analabha Roy}
\email[Email:]{daneel@utexas.edu}
\affiliation{Department of Physics, The University of Burdwan, Golapbag, Bardhaman - 713 104, India}

\begin{abstract}
	Dynamical many-body freezing occurs in periodic transverse field-driven integrable quantum spin systems. Under freezing conditions, quantum dynamics causes practically infinite hysteresis in the drive response, maintaining its starting value. We find similar resonant freezing in the Lipkin-Meshkov-Glick (LMG) model. In the LMG, the freezing conditions in the driving field suppresses the heating postulated by the \textit{eigenstate thermalization hypothesis} (ETH)		
	by inducing \textit{dynamical many-body localization}, or DMBL. This is in contrast to  Many Body Localization (MBL), which requires disorder to suppress ETH. DMBL has been validated by the inverse participation ratio (IPR) of the quasistationary Floquet modes. Similarly to the TFIM, the LMG exhibits high-frequency localization only at freezing points. IPR localization in the LMG deteriorates with an inverse system size law at lower frequencies, which indicates heating to infinite temperature. Furthermore, adiabatically increasing frequency and amplitude from low values raises the Floquet state IPR in the LMG from nearly zero to unity, indicating a phase crossover. This occurrence enables a future technique to construct an MBL engine in clean systems that can be cycled by adjusting drive parameters only.
\end{abstract}

\maketitle

\section{\label{sec:introduction} Introduction}
In the past few years, periodically driven quantum many-body systems have been of considerable theoretical and experimental interest~\cite{bordia_periodically_2017, sahoo_periodically_2019}. Under certain conditions in the drive parameters, they can experience dynamical many-body freezing (DMF), which causes the response to freeze completely to its initial value at all times~\cite{das_exotic_2010, sirshendu:freezing, asmi:floquet:thermalization}. This arises as a consequence of additional approximate symmetries that occur at the freezing points~\cite{asmi:scars}. DMF has been demonstrated via the rotating wave approximation (RWA) in the driven transverse field Ising model(TFIM) with nearest-neighbor interactions ~\cite{mbeng_quantum_2020} and is shown to be protected when translational invariance is explicitly broken (say, by disorder) ~\cite{yamada_localization_2022, roy_fate_2015}. 

The utilization of Floquet theory simplifies the analysis of time-periodic systems. For closed quantum systems governed by the time-dependent Schr\"odinger equation, the \textit{Floquet Hamiltonian} allows for a mapping of the time-dependent dynamics into the dynamics of 	
a time-independent effective Hamiltonian, provided the system is strobed at integer multiples of the time period of the drive. The time independent eigenstates of the effective Hamiltonian correspond to quasistationary \textit{Floquet Modes} of the original Hamiltonian. The temporal progression of the system comes from  phase coefficients that capture the dynamics \cite{li_floquet_2018,eckardt_high_frequency_2015}.	

Any sufficiently complex nonintegrable many-body system is expected to thermalize according to the eigenstate thermalization hypothesis (ETH) despite the fact that closed quantum dynamics preserves the memory of the initial state of the system. This arises due to the properties of the matrix elements of observables 
in typical states\cite{zhang_floquet_2016}. The ETH can be readily adapted to time-periodic systems using Floquet theory (the Floquet-ETH, or FETH \cite{Mori_2018, Kim_2014, Mizuta_2020, Mori_2023_1}). Nonetheless, the conditions for ETH to hold are not particularly strong, and the density matrix of the system can fail to approach one that is described by a thermal expression.  Thermal systems must conduct because they exchange energy and particles internally during thermalization. Thus, insulating systems can be naturally athermal;  many-body localization (MBL) is a well-studied case~\cite{khemani_phase_2016}. This phenomenon is stable against local perturbations, and constitutes an exotic state of matter with far-reaching implications in theoretical physics, as well as in practical applications\cite{yunger_halpern_quantum_2019}.

The addition of disorder has been identified as a crucial component in the onset of MBL. In that case, thermalization is prevented by disorder-induced localization. Nonetheless, alternative approaches to MBL in strongly interacting disorder-free systems~\cite{diptiman2014, Carleo2012,aditya2023dynamical}, inhomogeneous systems~\cite{alessandro_markus, Grover2014,miles2015,Smith2017}, and by inducing disorder in the emergent physics~\cite{MBL_emergent_disorder} and by other effective means ~\cite{miles2015} (albeit with strong finite-size effects), have been reported. An alternative approach to realizing MBL in disorder-free \emph{homogeneous} many-body systems involve \textit{Floquet Engineering}, where a time-periodic drive is introduced, and the drive parameters tuned to introduce a clustering of quasistationary energies in a manner similar to localization\cite{zhang_floquet_2016}.

In this article, we use the fact that emergent approximate symmetries can be engineered in Floquet systems~\cite{Engelhardt2013,asmi:scars} and apply it to long-range interactions. This results in \textit{dynamical many-body localization} (DMBL) at resonant values of the drive parameters, and complete thermal behavior at other values. This phenomenon is distinct from DMF in the TFIM, since clean TFIM systems, being integrable, never thermalize. To demonstrate the onset of MBL, we investigate the driven Lipkin-Meshkov-Glick (LMG) model\cite{lmg1965_1,lmg1965_2, lmg1965_3, debergh_2001, ribeiro2008,Engelhardt2013,titum2020}, a long-range system that extends the nearest-neighbor interactions in the TFIM to  all-to-all interactions~\cite{campa_statistical_2009, eisele_multiple_1988, canning_class_1992}. Previous works on driven LMG models have either focused on the low-frequency regime where FETH is applicable~\cite{russomanno_thermalization_2015}, or the onset of localization inferred from observable dynamics~\cite{Engelhardt2013, lmg:fidelity, Russomanno2017}.  We have recovered the onset of DMBL in this system by inspecting the values of the high-frequency inverse participation ratio (IPR) of the Floquet modes themselves, a result that is valid for infinite times, thus being a better indicator of localization.

In addition, we compare the degree of localization of the quasistationary Floquet modes in the LMG model with the TFIM, as well as other short-range models with lower symmetry. To do so, we look at the IPR of the Floquet modes in the representation given by the eigenstates of the symmetry-breaking field. The IPR, closely related to the concept of quantum purity, is defined as the formal sum of the square of the density in some physically meaningful space or representation. A high IPR of a stationary state denotes low participation in most of the representation, and a low IPR distributes participation uniformly across the representation, leading to ergodic dynamics\cite{vu_fermionic_2022}. Thus, IPR~\cite{Misguich2016} is a useful tool for witnessing MBL of a quantum system. For an MBL system, the IPR is unity, and it scales inversely with the {number of spins} when it is thermally distributed~\cite{calixto_inverse_2015}.

In the first section of this paper, we present all essential theoretical frameworks. Our results for the LMG model are presented next in Sec.\ref{sec:level3}. In that section, we have used the RWA ~\cite{fujii_introduction_2017}, where only the slowest rotating terms in the Fourier expansion of the Hamiltonian in a frame corotating with the symmetry breaking drive field are retained. In addition, we have the obtained {numerical simulations of} the Floquet modes and their IPR. They are used to probe the system dynamics in the high and low-frequency domains at both limits of $\beta$. In Sec. \ref{sec:level4}. we have used phase space plots to contrast the low and high frequency limits of the LMG model in the thermodynamic limit by mapping it to an equivalent classical Hamiltonian system. Finally, in Sec. \ref{sec:level5}., we have looked at numerical computations of the IPR of the Floquet modes for different values of the drive parameters, well beyond those that allow for the RWA. We achieved a smooth crossover from a thermal phase to a DMBL phase by increasing the frequency adiabatically while remaining at a freezing point. However, the DMBL phase is unstable at any finite driving frequency in the thermodynamic limit. This result is in contrast to the behavior of short-range models reported in the literature, in which a nonanalytic transition from a thermal phase to a strictly local phase takes place at finite driving parameters~\cite{asmi:floquet:thermalization}. Finally, we conclude with discussions and outlook.

\section{\label{sec:background} Background}

The ETH is a series of conjectures that allows for the thermalization of an isolated quantum many-body system. The state of the system, $\ket{\psi(t)}$, evolves according to the Schr\"odinger equation $\hat{H}\ket{\psi(t)} = i\displaystyle\frac{\partial}{\partial t}\ket{\psi}$. The Hamiltonian $\hat{H}$ is assumed to be \textit{nonintegrable}, in that  {it lacks an extensive number of conserved quantities that can be written as a sum of local operators, that is to say, there are no set of observables   $\hat{O}_s = \sum_i \hat{L}_i$ such that $\comm{\hat{O}_s}{\hat{H}}=0$. Here, the $\hat{O}_s$ constitute an arbitrary CSCO (complete set of commuting observables), and $\hat{L}_i$ are \textit{local}, each having subextensive support in the system}~\cite{Sutherland2004}. In addition, we postulate an ``equilibrium'' value $A_{eq}$ for every observable $\hat{A}$, such that
\begin{equation}
	\label{eq:aeq}
	A_{eq}(E)\equiv \frac{\Tr \left(\hat{A}e^{-\beta \hat{H}}\right)}{\Tr(e^{-\beta \hat{H}})},
\end{equation}
where $E=\expval{\hat{H}}{\psi(t)}$ is the conserved energy of the system, and $\beta = 1/(k_B T)$ is the inverse temperature, and $k_B$ is the Boltzmann constant.

To put it simply, ETH proposes that this many-body Hamiltonian undergoes thermalization as seen in the \textit{long-time averages} of observables, with the eigenstates bearing resemblance to thermal states. The aforementioned hypothesis serves as a valuable instrument for comprehending the conduct of stimulated quantum systems and their correlation with thermal equilibrium. This assertion can be justified by examining the expectation value of an observable $\hat{A}$ as it evolves under the Schr\"odinger equation. To see this, we first expand the state of the system $\ket{\psi(t)}$ as:
\begin{equation*}
	\ket{\psi(t)} =  \sum_m c_m(t)\ket{m(0)},\\ 
\end{equation*}
where $\ket{m(0)}$ represents the eigenstates of $\hat{H}(0)$ with energy $E_m$. The coefficients $c_m (t)$ describe the time-dependent amplitude of the expansion.
Plugging these expansions into the expression for the expectation value, we obtain the long-time average of the expectation value~\cite{abanin_colloquium_2019}:
\begin{equation}
	\overline{\expval{\hat{A}(t)}} 
	= \sum_{m,k} \overline{c_m^\ast(t) c_k(t)}
	\mel{m(0)}{\hat{A}}{k(0)},
\end{equation}
where the overline indicates the following operation for any time-dependent quantity $\mathcal{O}(t)$, 
\begin{equation}
	\overline{{\mathcal{O}}} \equiv \lim_{t\rightarrow\infty}\frac{1}{t}\int^t_0\;\mathrm{d}\tau\; {\mathcal{O}(\tau)} .
	\label{eq:lt_avg}
\end{equation}
The matrix elements $\mel{m(0)}{\hat{A}}{k(0)}$ are said to satisfy the Srednicki ansatz~\cite{Srednicki1994,Srednicki_1999}:
\begin{multline}
	\mel{m(0)}{\hat{A}}{k(0)} \approx A_{eq}\left(\frac{E_m+E_k}{2}\right) \delta_{mk} +\\ e^{-\frac12 S\left(\frac{E_m+E_k}{2}\right)} f\;\left(\frac{E_m+E_k}{2}, E_m-E_k\right)R_{mk}.
\end{multline}
Here, $S$ is the thermodynamic entropy and $R_{mk}$ are elements of a random matrix with vanishing mean and unit variance. What this means for the ensuing dynamics is that the system explores the accessible Hilbert space uniformly,  and the matrix elements $\bra{m(0)}\hat{A}(t)\ket{k(0)}$ become indistinguishable for most pairs of $m$ and $k$.
Applying this ansatz and taking the thermodynamic limit by ignoring terms $\mathcal{O}(e^{-S/2})$, the expression for the expectation value becomes:
\begin{multline*}
	\overline{\expval{\hat{A}(t)}} \approx \sum_m \overline{\abs{c_m(t)}^2}\; {A}_{eq}\left(E_m\right)\\
	\approx {A}_{eq}(E)\sum_m \overline{\abs{c_m(t)}^2} = A_{eq}(E),
\end{multline*}
where, in the last step, we utilized the fact that $A_{eq}$ is a smooth function, and that the states with energies far from $E$ have $\abs{c_m(t)}^2\approx 0$. Therefore, in the limit of large systems the expectation value of an observable $\hat{A}$ is approximately equal to the thermal expectation value $A_{eq}$. This is the essence of the ETH, which suggests that individual eigenstates of a quantum system can be described by statistical mechanics in the long-time limit.

We now generalize the ETH to nonintegrable many body systems that are closed, but not isolated. In that case, it is possible to impart a periodic time-dependence on the Hamiltonian while still ensuring unitary evolution. If the time period of the drive is $T$, and the corresponding drive frequency $\omega\equiv 2\pi/T$, then the Floquet theorem states that the solutions to the Schr\"{o}dinger equation can be written as $\ket{\psi(t)} = e^{-i\epsilon t/\hbar} \ket{\phi(t)}$, where the $\ket{\phi(t)}$ are $T$-periodic states called \textit{Floquet Modes}, the corresponding $\epsilon\in \mathbb{R}$, are called \textit{quasienergies}. Quasienergy values are not unique, and can be made to be bounded within a Floquet {Brillouin zone},\textit{viz.} a range $[-\omega/2, \omega/2]$\cite{holthaus_floquet_2016,vogl_effective_2020}. As a consequence, the unitary evolution operator can be spit into two parts as follows~\cite{Bukov2014}:
\begin{equation}
	\label{eq:propagator}
	\hat{U}(t) = e^{-i\hat{K}_F(t)}\;e^{-i\hat{H}_Ft}.
\end{equation}
Here, the micromotion operator $\hat{K}_F(t)$ is time-periodic in $T$, with $\hat{K}_F(0)=0$, and the Floquet Hamiltonian  {$\hat{H}_F = e^{i\hat{K}_F(t)} \left[\hat{H}(t)-i\partial_t\right] e^{-i \hat{K}_F(t)}$}. Thus, if the system is strobed at integer multiples of $T$ only, then the unitary evolution matches that of a time independent Hamiltonian $\hat{H}_F$. This can capture most of the exact dynamics at large frequencies. In such systems, the \textit{Floquet eigenstate thermalization hypothesis} (FETH)~\cite{Mori_2018, Mori_2023_1} posits that, subject to specific conditions and in the context of a system of significant size, the Floquet modes themselves exhibit thermal state-like behavior, \textit{i.e.}, $\hat{H}\approx \hat{H}_F$ in Eq.~(\ref{eq:aeq}). However, in contrast to the isolated systems, the loss of energy conservation allows for the mixing of all Floquet modes in the ensuing dynamics, not just those with quasienergies near $E$. Were this to actually happen in the ensuing dynamics,  it can be reconciled with ETH by ensuring that the right-hand side of Eq.~(\ref{eq:aeq}) is independent of $\beta$, \textit{i.e.}{, an infinite temperature ensemble}~\cite{alessio}. In other words, the nonequilibrium steady state of the system tends to an infinite temperature, maximum entropy density matrix.

However, drive parameters like amplitude, frequency, and duty-cycle strongly affect the structure of the Floquet modes $\ket{\phi}$. Thus, they can be engineered to prevent the kind of full mixing that would lead to infinite temperatures, manifesting suppression of thermalization dynamically. Thus, this type of \textit{Floquet Engineering} can produce \textit{dynamical many-body localization} (DMBL), where the system fails to reach thermal equilibrium and remains localized, possibly  near its initial state, even at large times. This paradigm seems similar to standard many-body localization\cite{Sougata2023, sierant_2023}, where disorder, locality, and integrability can cause athermality via breakdown in the Srednicki ansatz. However, DMBL {stems from periodic driving}, and thus can occur regardless of disorder, locality of observables, or system integrability, all of which have been studied for MBL onset~\cite{Sougata2023,Fabien2018,garratt_resonant_2022}.

Integrable many-body systems do not exhibit thermalization. When subjected to time-periodic drives, Floquet engineering allows for the introduction of additional approximate conserved quantities that dynamically suppress the evolution of certain observables by hysteresis. This type of \textit{freezing} of response
has been shown in integrable systems~\cite{roy_fate_2015}.
A paradigmatic example is  the driven TFIM in one dimension~\cite{stinchcombe_ising_1973}. The Hamiltonian is given by
\begin{align}
	\label{eq:tfim:hamilt}
	\hat{H}(t) &= \hat{H}_0 + h_z(t) \; \hat{H}_1,\\
	\hat{H}_0 &= -\frac{1}{2}\sum^N_{i=1} \hat{\sigma}^x_i \hat{\sigma}^x_{i+1},\\
	\hat{H}_1 &= -\frac{1}{2}\sum_{i=1}^N \hat{\sigma}^z_{i}.
\end{align}
Here, the undriven Hamiltonian $\hat{H}_0$ consists of nearest-neighbor interactions between $N$ number of  spin-$1/2$ particles on a one-dimensional spin network. The transverse field is denoted by $\hat{H}_1$, and is being varied by a time-periodic and harmonic signal $h_z(t) = h_0 + h\cos{\omega t}$, yielding a time period $T=2\pi/\omega$ with amplitude $h$, drive frequency $\omega$, and d.c. field $h_0$. This Hamiltonian can be readily transformed into a spinless pseudo-fermionic system via the Jordan-Wigner transformation~\cite{mbeng_quantum_2020}. When written in momentum space spanned by spinors $\psi_k = (c_{-k}, c^\dagger_k)^T$ of fermions at momentum $k$ created (annihilated) by operators $c^\dagger_k$ ($c_k$), the effective Hamiltonian
\begin{equation}
	\label{eq:TFIM:fermions}
	\hat{H}(t) = \sum_{(k,-k)-\mbox{pairs}} \psi^\dagger_k
	\Bigg[\ \bigg(f_k - h_z(t)\bigg)\tau_z +  \tau_x \Delta_k\Bigg]\psi_k ,
\end{equation}
where $f_k = \cos{k}$, $\Delta_k = \sin{k}$, $\tau_{xyz}$ are the three Pauli Matrices, and the sum is over distinct $(k, -k)$ \textit{Cooper Pairs}. We can transform our system to a frame that rotates with the time-varying symmetry-breaking field. This is achieved by the means of the unitary transformation operator~\cite{Engelhardt2013}
\begin{align}
	\label{eq:rotation:tfim}
	\hat{U}(t) &= \prod_k \hat{U}_k(t)\\
	\hat{U}_k(t) &= \exp{\Big[\frac{i h}{\omega}\sin{\omega t}\Big]\tau_z}.\nonumber
\end{align} 
The resulting transformed Hamiltonian $\hat{H}^\prime(t) = \hat{U}^\dagger(t)\;\hat{H}(t)\;\hat{U}(t)-i\hat{U}^\dagger(t)\;\partial_t\hat{U}(t)$ simplifies to
\begin{multline}
	\label{eq:rotated:tfim}
	\hat{H}^\prime(t) = \sum_{(k,-k)-\mbox{pairs}} \psi^\dagger_k
	\bigg[\ \tau_z f_k + \tau_x \cos{\big(\eta\sin{\omega t}\big)}  \\
	+ \tau_y \sin{\big(\eta\sin{\omega t}\big)}\bigg]\psi_k,
\end{multline}
where we defined $\eta=2h/\omega$. Using the Jacobi-Anger formula~\cite{arfkenmath}
\begin{equation}
	\label{eq:jacobi}
	e^{i \eta \sin{\omega t}} = \displaystyle\sum_{n=-\infty}^{\infty} J_n(\eta)\, e^{i n \omega t},
\end{equation} 
where $J_n(\eta)$ are Bessel Functions, the transformed Hamiltonian simplifies to \\
\begin{align}
	\hat{H}^\prime(t) =& \sum_{\substack{(k,-k) \\ \mbox{pairs}}} \psi^\dagger_k
	\bigg\{\ \tau_z f_k + 2\tau_x \Delta_k \sum_{n\geq 0} J_{2n}(\eta)\cos{\big(2n\omega t\big)} \nonumber\\
	&\hskip 0.7cm- 2\tau_y\Delta_k \sum_{n\geq 0} J_{2n+1}(\eta)\sin{\big[\left(2n+1\right)\omega t\big]}   \bigg\}\psi_k.
\end{align}
In the frequency regime  $\omega \gg f_k$, the long-time average $\hat{H}^{\mathrm{RWA}}\equiv\displaystyle\lim_{n\rightarrow\infty}\frac{1}{nT}\int^{nT}_0\mathrm{d}t\;\hat{H}^\prime(t)$ can serve as a suitable approximation for $H^\prime(t)$. This approximation, known as the \emph{rotated wave approximation} (RWA), eliminates the oscillating modes and results in an effective Hamiltonian that is independent of time,
\begin{equation}
	\label{eq:hrwa:tfim}
	\hat{H}^{\mathrm{RWA}} = \sum_{(k,-k)-\mbox{pairs}} \psi^\dagger_k
	\bigg[\ f_k\tau_z + 2 J_0(\eta) \Delta_k\tau_x \bigg]\psi_k.
\end{equation}
It is evident that by manipulating the drive parameters, specifically the amplitude denoted by $h$ and the frequency denoted by $\omega$, in a manner such that $\eta$ is positioned on a root of $J_0(\eta)$, the fermion number can be conserved to a significant extent at this particular point. Consequently, it is feasible to exercise direct control over $\hat{H}^{\mathrm{RWA}}$, resulting in a comprehensive suppression of the dynamics of otherwise responsive observables.

However, this phenomenon is not trivial to generalize, since the addition of integrability-breaking terms, such as longitudinal fields or additional spin-spin interactions, kills the suppression of fermion number dynamics.  For instance, if we apply an integrability-breaking longitudinal field $\hat{S}_x = \frac12 \sum_i \hat{\sigma}^x_i$, then the Hamiltonian becomes 
\begin{align}
	\hat{H}_{_\mathrm{TFIM+S_{x}}}(t) &=\frac12 \Bigg[\sum^N_{i=1}  \hat{\sigma}_{i}^{x} \hat{\sigma}_{i+1}^{x}+h \cos (\omega t) \sum^N_{i=1} \hat{\sigma}_{i}^{z}\nonumber\\
	&\hskip4cm  +\sum^N_{i=1} \hat{\sigma}_{i}^{x}\Bigg].
	\label{eq:tfim_sx}
\end{align}	
Performing the unitary transformation and taking the RWA yields a new RWA Hamiltonian
\begin{equation}
	\hat{H}_{_\mathrm{TFIM+S_{x}}}^{\mathrm{RWA}}= \hat{H}^{\mathrm{RWA}}+\frac12 J_{0}\left(\frac{h}{\omega}\right) \sum_i\hat{\sigma}^x_i.
	\label{eq:tfim_sx1}
\end{equation}
At the root of $J_0\left(\frac{2h}{\omega}\right)$, the longitudinal field survives. However, note that the Bessel function has asymptotic form $J_0\left(\frac{2h}{\omega}\right)\sim \left(\frac{2h}{\omega}\right)^{-1/2}\cos\left(\frac{2h}{\omega}-\frac{\pi}{4}\right)$, a good approximation for sufficiently  large $h/\omega$. In that limit, if $2h/\omega$ is chosen to lie at a root, then $J_0\left(\frac{h}{\omega}\right) \sim h^{-1/2}$, which is small for sufficiently large $h$. Thus, the contribution of the longitudinal field is substantially reduced if $h\gg\omega\gg1$, $J_0\left(\frac{2h}{\omega}\right)=0$, partially recovering dynamical freezing.
\begin{figure}[t!]
	\centering
	\includegraphics[width =8cm]{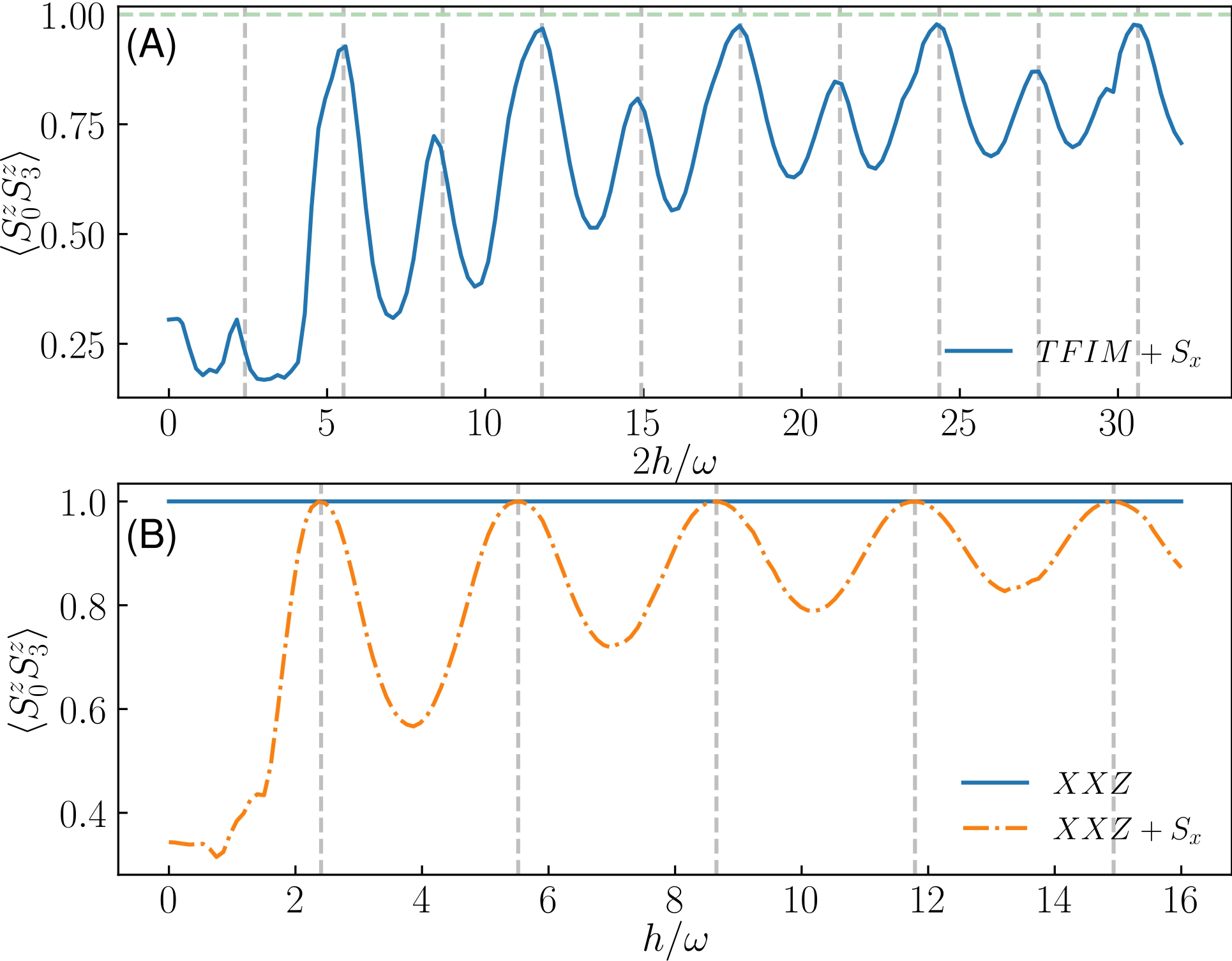}
	\caption{Time-average of spin-correlation $\expval{S^z_0 S^z_3}$ for a driven 1D spin-$1/2$ chain with periodic boundary conditions. The chain consists of $N=8$ spins, and was driven harmonically at fixed frequency $\omega=90$.  The system was initially populated in a fully $z-$polarized spin state and evolved numerically till time $t=300T$, where $T=2\pi/\omega$.  The vertical dashed lines represent the roots of the Bessel function $J_0$.  Panel (A) presents results for  the driven $\mathrm{TFIM+S_x}$ (the  Hamiltonian in Eq.~(\ref{eq:tfim_sx}), and panel (B) presents results for the driven  $\mathrm{XXZ}$ model (both with and without an $S_x-$field) from the Hamiltonian described in  Eq.~(\ref{eq:xxzsx}).
	For the former, correlations are suppressed at the lower roots of $J_0\left(2h/\omega\right)$, but approach unity at successively higher roots. For the latter, the correlations remain constant for all values of $h/\omega$. However, the introduction of an additional $S_x-$ field results in freezing only at the roots of $J_0\left(h/\omega\right)$.}
	\label{fig:ipr:tfimsx}
\end{figure}
This analysis can be supported by dynamical simulations for small-size systems. We have utilized QuTiP, the quantum toolbox in Python~\cite{qutip}, to numerically investigate the spin correlations $\expval{S^z_0 S^z_3}$ of a one-dimensional spin chain comprising of $N=8$ spins. The results are plotted against $2h/\omega$ for fixed $\omega$ in panel-A of  Fig.~\ref{fig:ipr:tfimsx}(A). At the lower roots of $J_0\left(2h/\omega\right)$,  the correlations are suppressed. However,  they  gradually increase towards unity when the drive parameters are adjusted to lie at progressively higher roots.

Interestingly, if we apply this approach to investigate the periodically driven $XXZ$ model, we notice that the system remains frozen for all drive parameters. The $XXZ$ Hamiltonian, including a longitudinal field, is given by
\begin{multline}
	\hat{H}_{_{XXZ+S_{x}}} = \frac12 \sum_{i=1} \bigg[ J \hat{\sigma}^x_i \hat{\sigma}^x_{i+1} +J  \hat{\sigma}^y_i \hat{\sigma}^y_{i+1}\\ + \Delta  \hat{\sigma}^z_i \hat{\sigma}^z_{i+1} + h\cos(\omega t)  \hat{\sigma}^z_i + \hat{\sigma}^x_i\bigg],
	\label{eq:xxzsx}
\end{multline}
Transforming to the moving frame  yields
\begin{equation}
\hat{H}_{_{XXZ+S_{x}}}^{mov} =\hat{H}_{_{XXZ}}^{mov} +  \frac12 \sum_i\Big[ \hat{\sigma}^x_i \cos(\zeta) + \hat{\sigma}^y_i \sin(\zeta)\Big].
\label{eq:xxzsxmov}
\end{equation}
Here, $\zeta \equiv \frac{h}{\omega}\sin(\omega t)$, and the moving frame $XXZ$ Hamiltonian is
\begin{equation}
\hat{H}_{_{XXZ}}^{mov} =  \sum_i\Bigg[J\left(\hat{\sigma}^+_i \hat{\sigma}^-_{i+1} + \hat{\sigma}^-_i \hat{\sigma}^+_{i+1}\right) + \frac{\Delta}{2} \hat{\sigma}^z_i \hat{\sigma}^z_{i+1}\Bigg],
\end{equation}
where $\hat{\sigma}^\pm_i$ are spin ladder operators. Note that the moving frame Hamiltonian has no $\zeta-$dependence. If we populate the system initially in a  fully $z-$polarized state, then $H_{_{XXZ}}^{mov}$ does not contribute to the dynamics at all. The remaining longitudinal field  will simply add single particle dynamics to the state. If we now apply RWA in Eq.~(\ref{eq:xxzsxmov}) by smoothing out all harmonic oscillations from the Jacobi -Anger expansion, we get
\begin{equation}
\hat{H}_{_{XXZ+S_{x}}}^{_{RWA}} = \hat{H}_{_{XXZ}}^{mov} +  \frac12 J_0 \left(\frac{h}{\omega}\right)\sum_i \hat{\sigma}^x_i,
\label{eq:xxz_sx_rwa}
\end{equation}
Thus, freezing can occur in the $XXZ$ model in the absence of a longitudinal field regardless of the drive parameters. However when the $S_x-$ field is introduced, freezing can occur only when the drive parameters are controlled in such manner that ($h/\omega$) lies on any root of the Bessel function. We have supported this conclusion with QuTiP simulations~\cite{qutip} similar to those described in the previous paragraph. The resultant spin correlations are plotted against $h/\omega$ for fixed $\omega$ in Fig.~\ref{fig:ipr:tfimsx}(B). 

\begin{figure}[t!]
	\centering
	\includegraphics[width = 8.5cm]{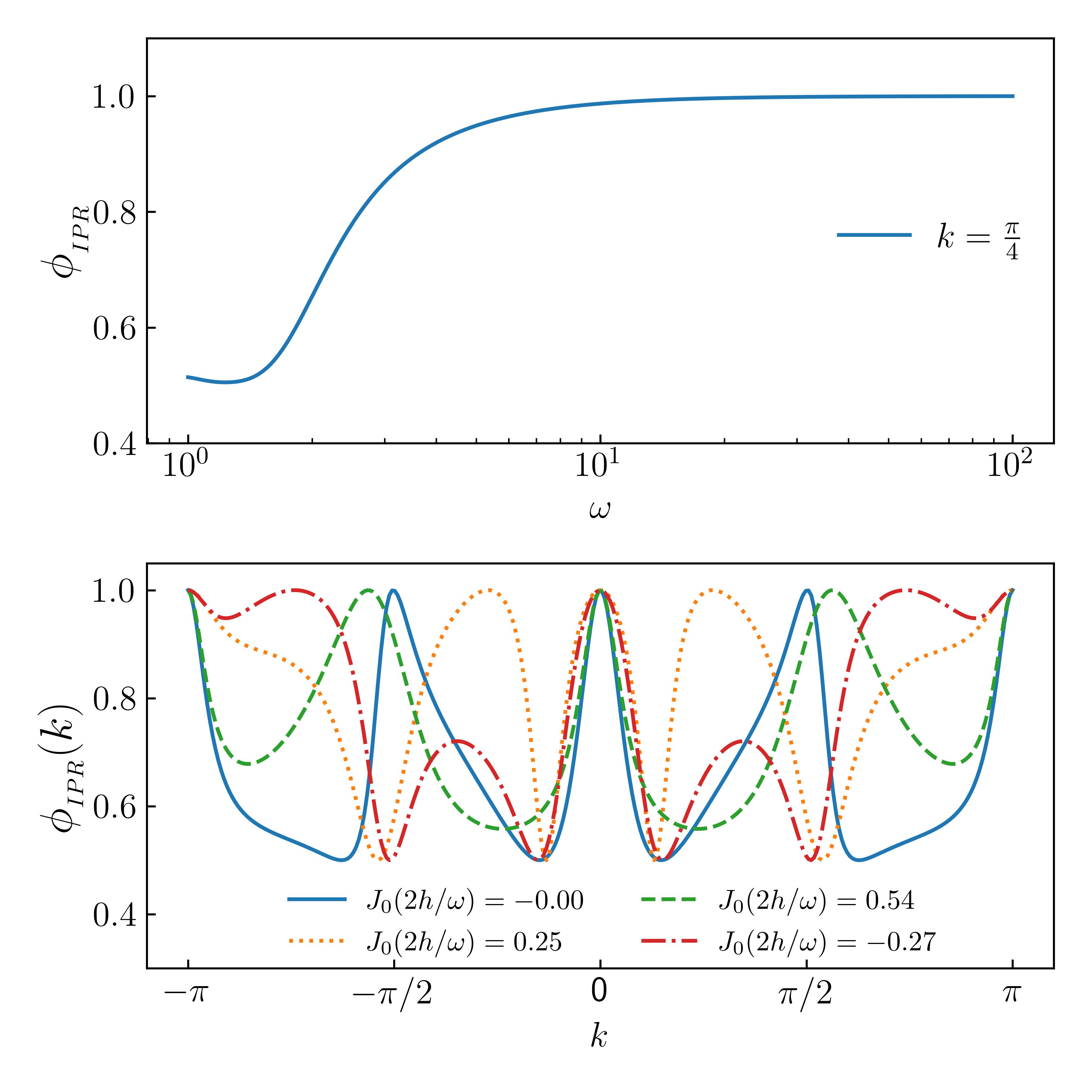}
	\caption{Reduced IPR obtained by adiabatically increasing $\omega$ (top panel $x$-coordinate) for one the floquet mode obtained from Eq.~(\ref{eq:ipr:ising}) at the root of $J_0(\eta)$ for N = 500. IPR is $\sim 0.5$ (localized yet not fully freezing) upto $\omega \sim 2$, after that, smoothly increased to unity (fully localized and freezing) at higher $\omega \geq 10$ (top panel, $y$-coordinate). At bottom panel cross-sections for four chosen amplitudes at $\omega =2$ are plotted for a brillouin zone ($x$-coordinate) with corresponding reduced IPRs ($y$-coordinate).}
	\label{fig:ipr:isinglowfrk}
\end{figure}

Now, if the integrability of the TFIM is broken by extending the range of $x-$spin interactions beyond nearest-neighbor, then freezing can \textit{still} be recovered through the emergence of additional approximate conserved quantities at the freezing point (see Ref.~\cite{rahaman2024time} for a detailed derivation). The case of all-to-all interactions is particularly interesting, as complete freezing can be achieved in that limit, as demonstrated in the next section. In such cases, freezing has the additional effect of inducing DMBL, suppressing Floquet-thermalization to infinite temperatures. Numerical quantification of localization of a specific (quasi) stationary state in a physically significant representation can be achieved through the computation of the IPR.  {The IPR is generally defined as the formal sum over the square of the local density in a physically meaningful space.}~\cite{mukherjee_modulation-assisted_2015,lin_many-body_2018,murphy_generalized_2011, torres-herrera_self-averaging_2020} {In the single particle case,  the IPR, } for a state $\ket{\psi}$ can be written as
\begin{equation*}
	\phi_{\mathrm{IPR}}\equiv \int \mathrm{d}\mathbf{x}\;\vert\ip{\mathbf{x}}{\psi}\vert^4.
\end{equation*}
This definition can be {applied} to {obtain} the IPR of a state $|\phi\rangle$ in a representation given by {any single particle} basis $\ket{m}$ as 
\begin{equation}
	\label{eq:ipr:defn}
	\phi_{{_\mathrm{IPR}}} \equiv \sum_m\vert\ip{m}{\psi}\vert^4.
\end{equation}
The smallest value of the IPR corresponds to a fully de-localized state, $\psi(x)=1/\sqrt{N}$ for a system of size $N$ \cite{torres-herrera_self-averaging_2020,trivedi_can_2005}. Values of the IPR close to unity correspond to localized states \cite{Misguich2016}. For a periodically driven system, we look at the IPR of the quasi-stationary Floquet modes at $t=T$, where $t=2\pi/\omega$ for drive frequency $\omega$. In the TFIM model, Eq.~(\ref{eq:hrwa:tfim}) indicates that, when $J_0(\eta)=0$, the Floquet modes are approximately given by the fermionic Fock states, which have a trivially unit IPR in the representation of the eigenmodes of the transverse field $\hat{H}_1$ in Eq.~(\ref{eq:tfim:hamilt}). Here, a particular Floquet mode can be decomposed into a direct product of cooper-pair states as $|\phi\rangle = \prod_{k,-k}|\phi^n_k\rangle$. In the RWA limit and at freezing, $|\phi^n_k\rangle$ has values of $|0\rangle, |k,-k\rangle$ for two values of $n=0,1$ respectively. We define the reduced IPR of $|\phi^n_k\rangle\; \forall k$ to be
\begin{equation}
	\label{eq:ipr:ising}
	\phi^{(n)}_\mathrm{IPR}(k) = \abs{\ip{0}{\phi^n_k}}^4 + \abs{\ip{ +k, -k }{\phi^n_k}}^4,
\end{equation}
where $n=0,1$. The full many-body IPR can be obtained from the reduced IPR in Eq.~(\ref{eq:ipr:ising}) by a product over all momenta in the Brillouin zone, yielding
\begin{equation}
	\label{eq:ipr:manybody}
	\phi_\mathrm{IPR} = \prod_k \phi^{(n)}_\mathrm{IPR}(k).
\end{equation}
In the RWA limit and at freezzing , this quantity is unity, indicating very low participation and the onset of freezing.
\begin{figure}[t!]
	\centering
	\includegraphics[width = 8.5cm]{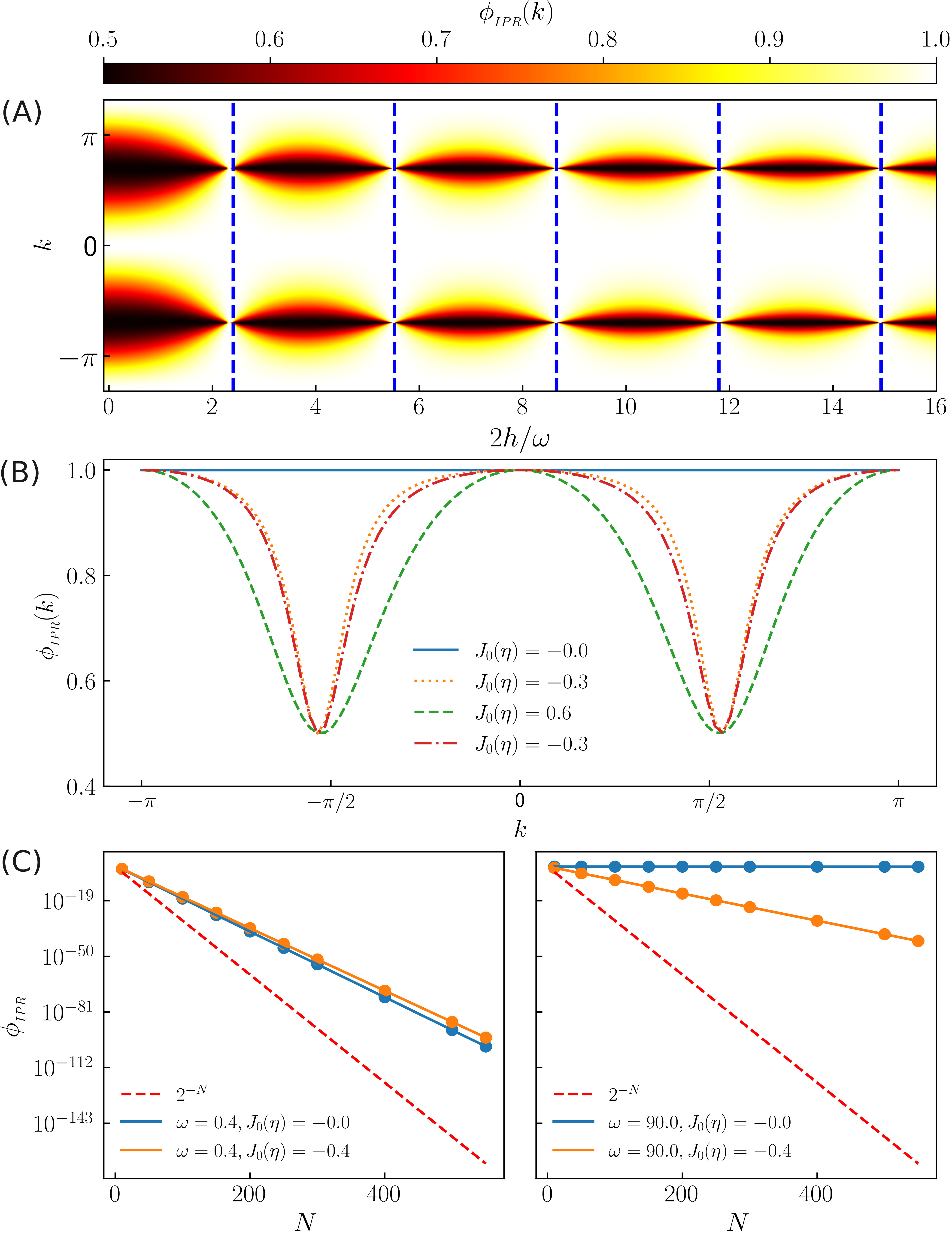}
	\caption{Reduced IPR [defined in Eq.~(\ref{eq:ipr:ising})] {for one of the two Floquet modes obtained from the exact dynamics of the TFIM for size} $N = 100, \omega=90$ {for the entire Brillouin zone [panel (A), y-coordinate] and a few drive amplitudes (top panel, x-coordinate). The dashed lines [in panel (A)] indicate the roots of} $J_0(\eta)${. The middle panel (B) shows cross-sections of the reduced IPR in} $k-$ {space for four chosen amplitudes. Finally, the bottom panels (C) show semi-log plots of the scaling with} $N$ of the full many body IPR as defined in Eq.~(\ref{eq:ipr:manybody}), with the left (C) panel for a small $\omega=0.4$, amplitude $h$ chosen to lie both in and out of the root of $J_0(\eta)$ {as indicated in the legend, with similar plots on the right(C) panel for a large} $\omega=90$. }
	\label{fig:ipr:tfim}
\end{figure}
Figure~\ref{fig:ipr:tfim} shows results from numerically simulating the TFIM dynamics. The reduced IPR for a particular Floquet mode recovered by simulating the exact Schr\"odinger dynamics over a single time period of the drive, and plotted as a function of momentum $k$ for different $\eta$'s. At freezing, when $\eta$ lies at the root of the Bessel function $J_0(\eta)$,  the {Reduced} IPR is {nearly} unity for all momenta. Consequently, so is $\phi_\mathrm{IPR}$.  Outside the freezing point, the IPR is unity only for some momenta because the effective Hamiltonian is perfectly diagonal at $k \in{\{-\pi, 0, \pi\}}$ as can be seen in the cross-sectional plots of Fig.~\ref{fig:ipr:tfim}. As we move away from the freezing point, the full IPR decreases exponentially in the system size, as can be seen in the bottom-right panel of Fig.~\ref{fig:ipr:tfim}.
However, as the TFIM is an integrable spin model, the IPR never drops to a value that is small enough to indicate thermalization. This can be seen in the bottom panels of Fig.~\ref{fig:ipr:isinglowfrk}, where the exponential decay of $\phi_\mathrm{IPR}(N)$ never approaches the thermodynamic scaling law $\phi_\mathrm{IPR}(N)\sim 2^{-N}$ for either small or large frequencies. Note that, at low frequencies, {complete localization} fails due to the unavailability of zero off-diagonal terms in the  effective transformed Hamiltonian, as well as the absence of integrability breaking terms to counteract the off diagonal terms. 
Because the TFIM can be mapped to a system of noninteracting particles as shown in Eq.~(\ref{eq:TFIM:fermions}), it is not appropriate to refer to the unit IPR region as "\emph{Many Body Localization}". The type of Floquet Engineering described above, however, can be easily applied to a broad class of nonintegrable systems where FETH is expected to hold in certain regions. Long-range spin systems, in particular, where the exchange energies between far-off spins are taken into account in the model Hamiltonian, are good candidates because they are known to thermalize when driven with low frequencies~\cite{russomanno_thermalization_2015}.
\section{\label{sec:level3}Long Range Interactions: The Lipkin Meshkov Glick Model: }	
The periodically driven {Lipkin Meshkov Glick (LMG)} model~\cite{lmg1965_1,defenu2018} for $N$ long-range spins is described by the Hamiltonian
\begin{equation}
	\hat{H}(t) = \hat{H}_0 + \big[h \cos{(\omega t)} + h_0\big]\; \hat{H}_1.
	\label{eq:driven:ham}
\end{equation}
Here, the undriven part $\hat{H}_0$ and the driven part $\hat{H}_1$ are, respectively, 
	\begin{eqnarray}
		\hat{H}_0 &=& -\frac{2}{N-1} \sum_{i<j}\hat{S}^z_i \hat{S}^z_j,\nonumber \\
		\hat{H}_1 &=& -2 \sum_i \hat{S}^x_i.
		\label{eq:h0h1}
	\end{eqnarray}
	The Kac-norm of $2/(N-1)$ arises from the choice to maintain the extensivity of the interaction energy. The Hamiltonian in Eq.~(\ref{eq:driven:ham}) commutes
with the total angular momentum $S^2 = \abs{\vec{S}}^2$, where $\vec{S_i}=\frac12 \sum_i \vec{\sigma}_i$. We now choose to populate the system in a state with $S^2=\displaystyle\frac{N}{2}\left(\frac{N}{2}+1\right)$. In that case, the dynamics remains invariant in the  $N+1-$dimensional space spanned by the common eigenstates of $P_{ij} \equiv \displaystyle\frac{1}{2}\left(1+ \vec{\sigma}_i\cdot\vec{\sigma}_j\right), \abs{S}^2$ and $S_z$; the so-called \textit{ totally symetric substace}, or TSS ~\cite{mori_prethermalization_2019}. Let the eigenvalues of $S^z$ in the TSS be $s_n$, and the eigenvectors be $\ket{s_n}$. Here, $s_n=-\frac{1}{2}+\frac{n}{N}$ and the index
$n= 0 (1) N$ has $N+1$ values. The dynamics is restricted to this invariant subspace, wherein the matrix elements of the Hamiltonian are given by
\begin{align}
	\mel{s_i}{\hat{H_0}}{s_j} &= -\frac{2}{N-1} s^2_i \delta_{ij},\nonumber\\
	\mel{s_i}{\hat{H_1}}{s_j} &= -\Bigg[\sqrt{\frac{N}{2}\left(\frac{N}{2}+1\right) - Ns_i(Ns_{i + 1})}\delta_{i+1, j} \nonumber\\ 
	&\hskip 0.7cm +\sqrt{\frac{N}{2}\left(\frac{N}{2}+1\right) - Ns_i(Ns_{i- 1})}\delta_{i-1,j}\Bigg].
	\label{eq:lmg:tssham}
\end{align}
These allow for a numerical representation of the Hamiltonian in the TSS.

Next, we transform the Hamiltonian to the rotated frame given by the operator
\begin{equation}
	\hat{U}(t)=\exp [i \frac{h}{\omega} \sin (\omega t) \hat{H}_{1}].
\end{equation}
This is analogous to the rotation performed for the TFIM in Eqs.~(\ref{eq:rotation:tfim}) and~(\ref{eq:rotated:tfim}). Defining $\tau = \displaystyle\frac{h}{\omega}\sin{\omega t}$, we use the fact that $\hat{H}_1 = 2 S^x$, as well as the following identity obtained by using the Baker-Campbell-Hausdorff formula,
\begin{equation}
	e^{i 2\tau\hat{S^{x}}} \hat{S^{z}} e^{-i 2\tau \hat{S^{x}}}=\hat{S^{z}} \cos \left(2\tau\right)+\hat{S}^{y} \sin (2\tau),
\end{equation}
to simplify the transformed Hamiltonian $\tilde{H}(t) = \hat{U}^\dagger(t)\;\hat{H}(t)\;\hat{U}(t) - \hat{U}^\dagger(t)\;\partial_t\hat{U}(t)$, yielding
\begin{align}
	\label{eq:lmg_rotated}
	\tilde{H}(t)& = -\frac{1}{N-1}\Big[\big(\hat{S}^z\big)^2 \big(1+\cos{4\tau}\big)+ \big(\hat{S}^y\big)^2 \big(1-\cos{4\tau}\big)\nonumber \\  
	&\hskip 2.5cm + \pb{\hat{S}^y}{\hat{S}^z}
	\sin{4\tau}\Big] - 2 h_0 \hat{S}^x.
\end{align}
Next, we define $\eta\equiv 4h/\omega$ and use the Jacobi-Anger formula in Eq.~(\ref{eq:jacobi})
to expand $\tilde{H}(t)$. This yields
\begin{widetext}
	\begin{multline}
		\label{eq:lmg_jacobiexp}
		\tilde{H}(t)= -\frac{\hat{S}^2}{N-1} +  \frac{\big(\hat{S}^x\big)^{2}}{N-1} - 2h_0 \hat{S}^x - \frac{J_0(\eta)}{N-1}\bigg[\big(\hat{S}^z\big)^{2} - \big(\hat{S}^y\big)^{2} \bigg] - \frac{2}{N-1}\;\Big[\big( \hat{S}^z\big)^2 - \big( \hat{S}^y\big)^2\Big]\;\sum^\infty_{k=1}\;J_{2k}(\eta)\cos{\big(2k\omega t\big)}\\
		- \frac{2}{N-1}\;\big\{ \hat{S}^y,  \hat{S}^z \big\}\;\sum^\infty_{k=1}J_{2k-1}(\eta)  \sin{\Big[\big(2k-1\big)\omega t\Big]}.
	\end{multline}
\end{widetext}
If $\omega$ is large enough to smooth out the harmonic components, then we obtain the RWA,
\begin{multline}
	\tilde{H}(t)\approx \tilde{H}_{\mathrm{RWA}}\equiv -\frac{\hat{S}^2}{N-1} +  \frac{\big(\hat{S}^x\big)^{2}}{N-1} - 2h_0 \hat{S}^x\\
	- \frac{J_0(\eta)}{N-1}\bigg[\big(\hat{S}^z\big)^{2} - \big(\hat{S}^y\big)^{2} \bigg].
	\label{eq:lmg_rwa}
\end{multline}
\begin{figure}[t!]
	\centering
	\includegraphics[width=9.3cm]{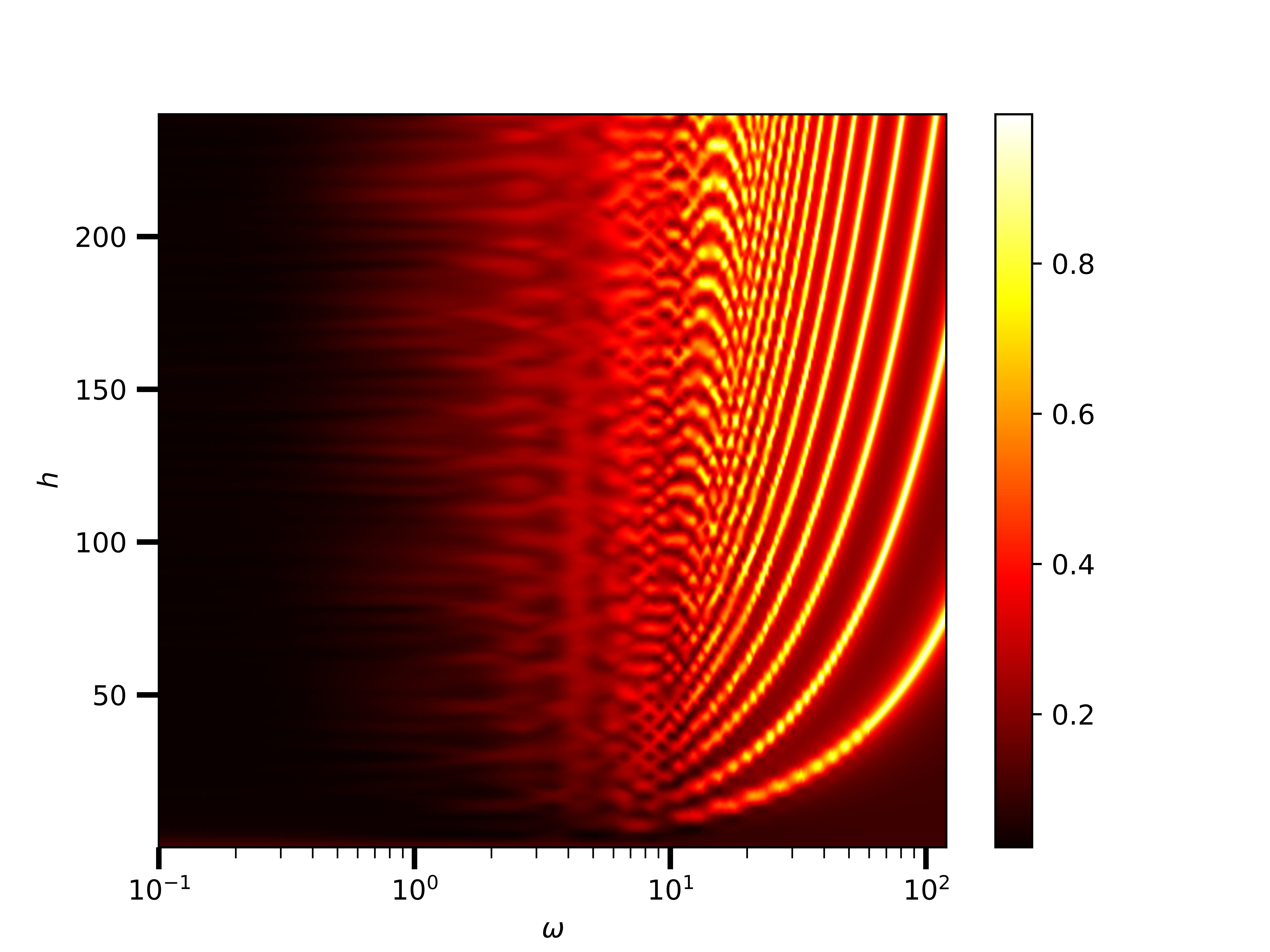}
	\caption{Plot of the numerically averaged IPR [IPR computed using Eq.~(\ref{eq:ipr:lmg})] in the TSS plotted in the $h$-$\omega$ plane for $N=100$ spins.To display the thermalized region more clearly, $\omega$ is plotted on a logarithmic scale on the $x$-coordinate. Note that, since the IPR is clearly nonnegative, an average IPR of zero means that all Floquet states have zero IPR. Furthermore, the boundedness of IPR in $\phi_\mathrm{IPR}(n) \leq 1$ ensures that if he average IPR is unity, then all Floquet states have unit IPR.}
	\label{fig:lmg_phasediag}
\end{figure}
If the drive amplitude $h$ is adjusted such that $\eta$ lies at a root of $J_0(\eta)$ (the localization point), the RWA Hamiltonian is diagonal in the representation of the simultaneous eigenstates of transverse field $\hat{S}^x$, and $S^2$, yielding an IPR of unity in that representation, similar to the TFIM in the previous section. Note however, that if the DC transverse field $h_0$ is set to $0$, then, at the localization point, the RWA Hamiltonian $\tilde{H}_{\mathrm{RWA}}\sim
\big(\hat{S}^x\big)^2$ in the TSS. The eigenvalues are two-fold degenerate. This produces infinitely many (Floquet) eigenmodes in the degenerate subspace whose IPRs may not always be unity in the $S^x$ representation. The removal of this degeneracy necessitates the inclusion of the d.c. field $h_0$. However, note that rational values of $h_0$
may add accidental degeneracies in $\tilde{H}_{\mathrm{RWA}}$. To see this, note that, at a localization point, the eigenvalues of $\tilde{H}_{\mathrm{RWA}}$ in the TSS are given by
\begin{equation}
	\rm{Eigs}\bigg[\tilde{H}_{\mathrm{RWA}}\bigg] = \frac{\big(\frac{N}{2}-m\big)^2}{N-1} - 2h_0 \bigg(\frac{N}{2}-m\bigg),
\end{equation}
where the half-integer $-N/2\leq m \leq N/2$ is the eigenvalue corresponding to a particular eigenstate $\ket{m}$ of the symmetry-breaking field $\hat{S}^x$. To ensure that no additional degeneracies occur, we have to set $h_0$ in  such a way that no two energies accidentally coincide.
\begin{figure}[t!]
	\centering
	\includegraphics[width = 8.5cm]{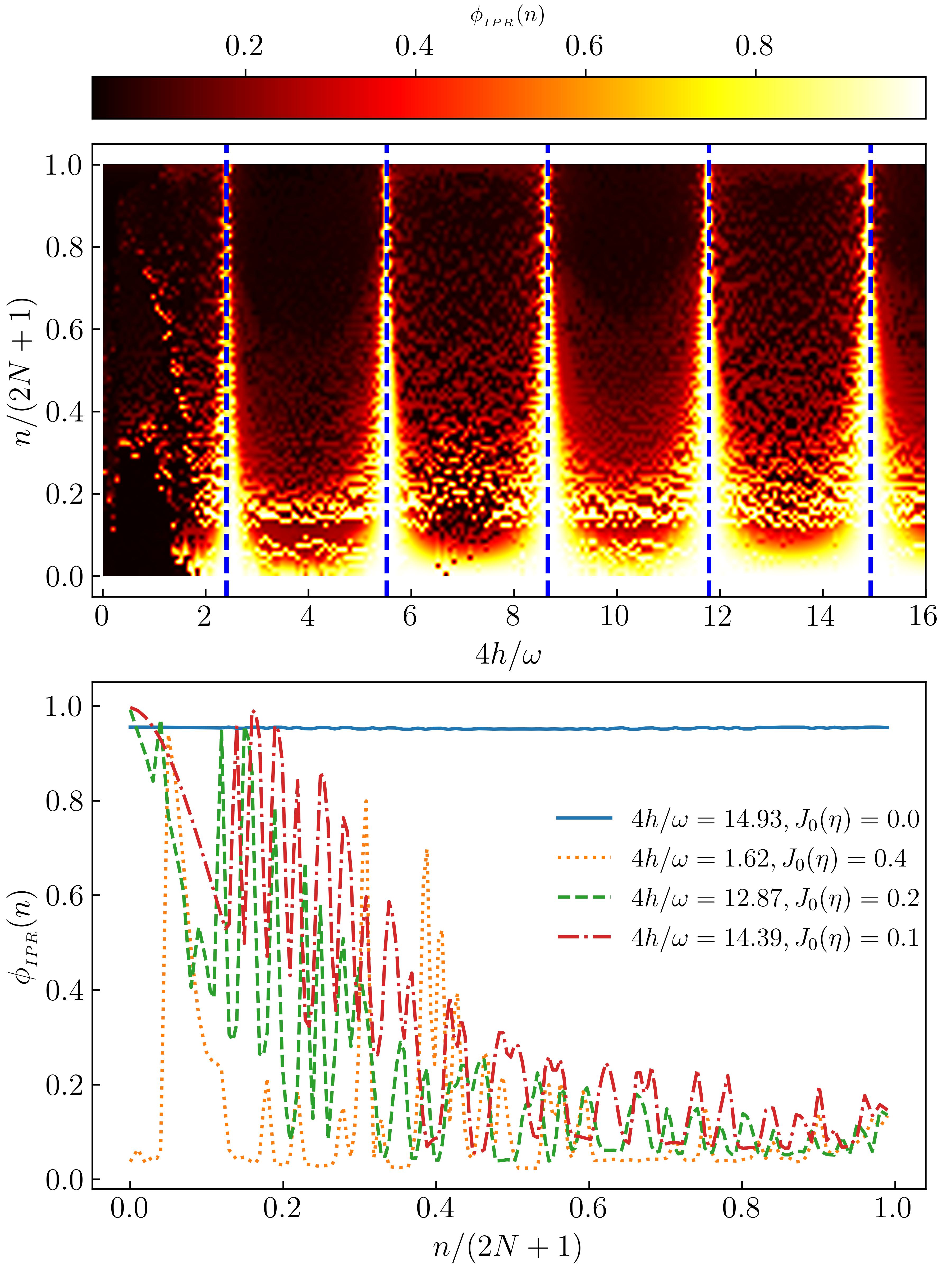}
	\caption{IPR density plot for all possible Floquet modes (top panel, $y$-coordinate) for different values of $\eta = 4h/\omega$ (top panel $x$-coordinate), deduced from Eq.~(\ref{eq:ipr:lmg}) for exact LMG Hamiltonian for N = 50. Blue dashed lines are roots of $J_0(\eta)$. At bottom panel cross-section of IPR ($y$-coordinate )for four different $\eta$'s plotted for all possible floquet modes(bottom panel, $x$-coordinate) at $\omega=90$. IPR founds to be $\sim$ unity for all Floquet modes at roots of $J_0$.}
	\label{fig:lmg_ipr_exact}
\end{figure}
If $N\gg 1$ (substantially large), then this condition can be readily met by assuring that $(1-2h_0)^{-1}$ is never an integer that is divisible by $N$. To ensure this in our numerical simulations, we have kept $h_0$ at a small irrational value.
The localization of the Floquet states at freezing is supported by exact numerical results, as can be seen in the phase diagram Fig.~\ref{fig:lmg_phasediag}. Here, we have plotted the arithmetic mean over all Floquet states of the IPR 
in the TSS for each point in the $h$-$\omega$ plane for $N=100$ spins.  The IPR in $S^x$ representation is
\begin{equation}
	\phi_\mathrm{IPR}(n) = \sum_m \abs{\ip{m}{\phi^n}}^4.
	\label{eq:ipr:lmg}
\end{equation}
As can be readily seen in the figure, the IPR is essentially zero when  $\omega \lesssim 1$. There is considerable structure in the phase diagram for larger drive frequencies, and along the lines given by the roots of $J_0(\eta)$, the IPR is essentially unity, in agreement with Eq.~(\ref{eq:lmg_rwa}).

In Fig.~\ref{fig:lmg_ipr_exact}, we show plots of the IPR of the Floquet modes $\ket{\phi^n}$ for $S^2 = \big(N/2\big)\big(N/2 + 1\big)$.
These plots were obtained numerically by diagonalizing the propagator $U(t)$ at $t=T$, where $U(t)$ is defined in Eq.~(\ref{eq:propagator}). This propagator was obtained from simulations of the exact quantum dynamics using QuTiP~\cite{qutip}. We kept the frequency at a fairly large value $\omega = 90$ where we expect that RWA would be valid, and $N=\mathcal{O}(10^2)$. The density plot in the upper panel of Fig.~\ref{fig:lmg_ipr_exact} depicts the IPR of the Floquet states; the abscissa ($x$-coordinate)  $\eta=4h/\omega$ and the ordinate is $n/(2N+1)$, where $n\leq 2N$ is a nonnegative integer that indexes the Floquet states in increasing order of $m$. The dashed vertical lines correspond to the roots of $J_0(\eta)$. Comparing with the IPR of the TFIM in Fig.~\ref{fig:ipr:tfim}, we can see a very similar pattern in the immediate neighborhood of the roots. Evidently, the IPR approaches a value of one for sufficiently large values of the roots, strongly suggesting full DMBL. Deviations occur at the smallest root of $J_0(\eta)$ (around $\eta = 2.405$) due to the contributions from higher order terms in Eq.~(\ref{eq:lmg_jacobiexp}). Thus, a higher root is favored for DMBL.

The bottom panel of Fig.~\ref{fig:lmg_ipr_exact} contains cross-sections of the full IPR plot for selected values of $\eta$ as indicated in the legend. When the drive amplitude $h$ is adjusted such that $\eta$ is close to a root of $J_0(\eta)$, the Floquet States are mixed, but not entirely thermal, since the IPR does not fall to $\mathcal{O}(N^{-1})$, indicating that localization persists to some extent. However, the further we are from the roots, the closer the IPR gets to one predicted by thermalization.

\begin{figure}[t!]
	\centering
	\includegraphics[width=9.3cm]{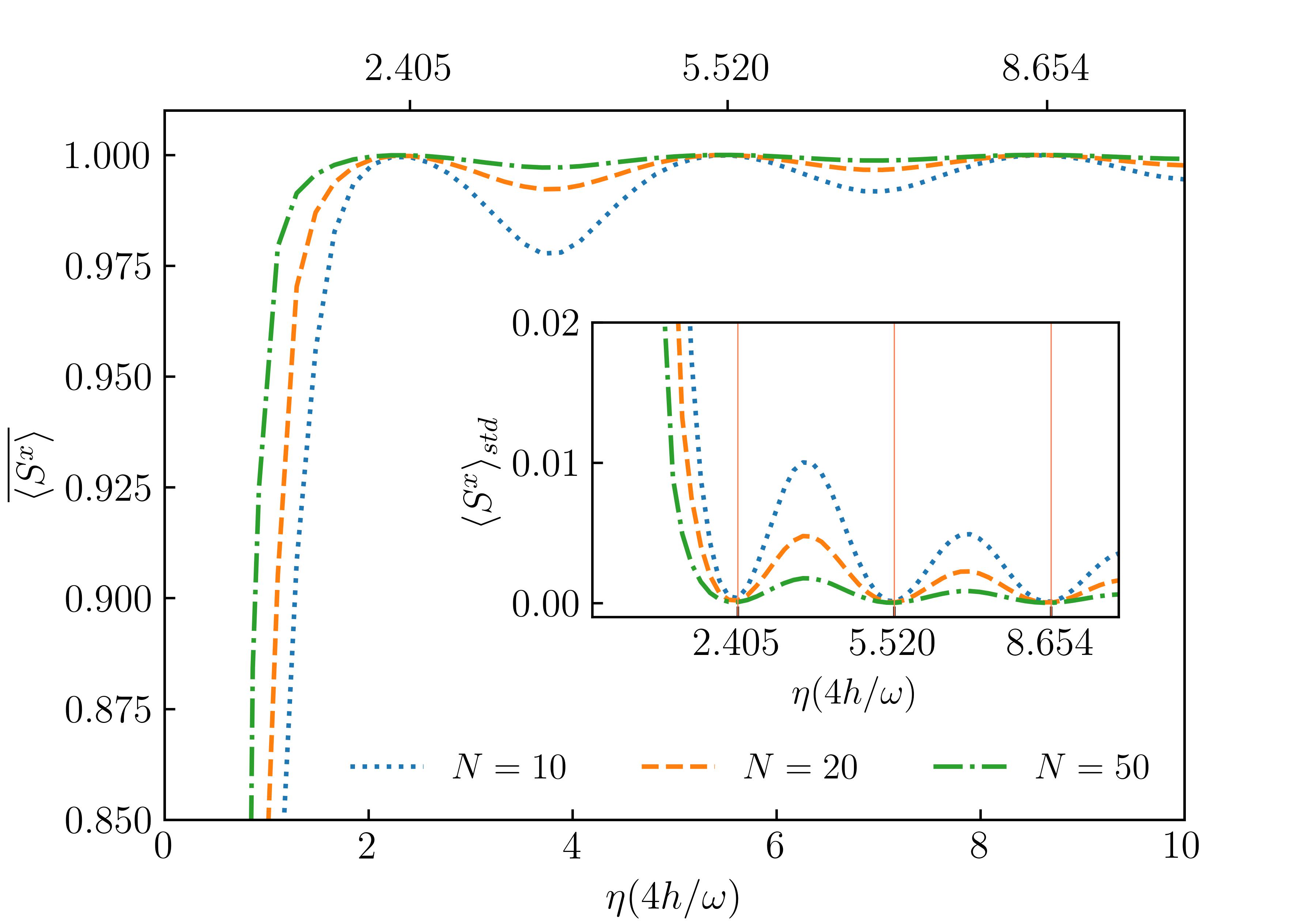}
	\caption{Temporal average of $\expval{\hat{S^x}}$ ($y$-coordinate) for different $\eta$'s ($x$-coordinate) is plotted for $\sim 200T$ at higher $\omega$ for different N=10, 20, 50. $\overline{\langle S^x \rangle}$ is found to be unity at roots of $J_0(\eta)$. At points away from these points, $\expval{\hat{S^x}}$ falls below unity. The corresponding standard deviation $\expval{\hat{S^x}}_{std}$ supports the variation of $\overline{\langle S^x \rangle}$ (inset figure). $\expval{\hat{S^x}}_{std}$ is $\sim 0$ describing a full freezing of the system at roots of $J_0(\eta)$ (red vertical solid lines).}
	\label{fig:sx_conserve}
\end{figure}

Figure~\ref{fig:sx_conserve} shows plots of the long-time average (from $t=0$-$200T$) of the field amplitude $\expval{\hat{S^x}}$ as a function of $\eta$. The system is started from the fully polarized state $s_n=N/2$ in the TSS and the dynamics simulated. The average is plotted for different values of amplitude $h$, keeping the frequency fixed at a high value of $\omega=90$. It is clearly very close to unity at roots of $J_0(\eta)$ and falls at points away from it, indicating that $S^x$ is approximately conserved at the localization points.

\begin{figure}[t!]
	\centering
	\includegraphics[width =8.5cm]{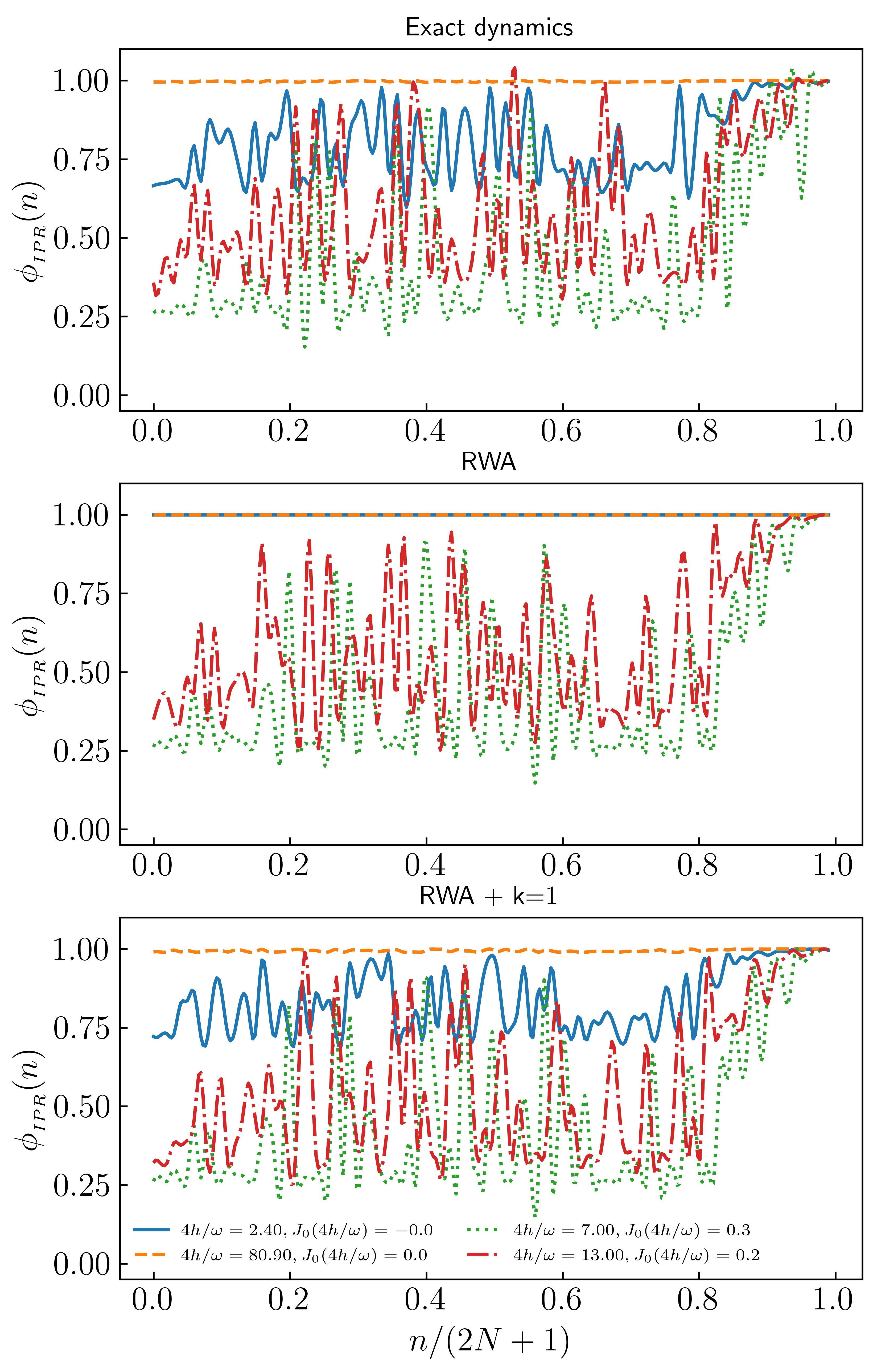}
	\caption{The comparison between IPR for exact dynamics and RWA with corresponding correction orders. IPR is calculated for four different $\eta$'s and corresponding $J_0(\eta)$ values for colors, Blue :$\eta = 2.40, J_0(\eta) = 0.0$, dashed orange:$\eta = 80.9, J_0(\eta) = 0.0$, Green:$\eta =7.0, J_0(\eta) = 0.3$, Red:$\eta = 13, J_0(\eta)= 0. 2$. At low root of $J_0(\eta)$ IPR is not unity (Blue curve) where at higher root (orange dashed) it is unity while at points away from roots IPR are less than unity in the exact (top panel) plot. RWA does not match with the exact plot. At all roots of $J_0(\eta)$ IPR is unity(middle panel). RWA with additional higher order terms exhibit similar system pattern(Bottom panel) with exact dynamics.}
	\label{fig:lmg_ipr_rwa11}
\end{figure}

Small deviations do occur due to the role of  higher order terms in the rotated Hamiltonian in Eq.~(\ref{eq:lmg_rotated}). This can be demonstrated quantitatively by comparing the IPR obtained from the exact dynamics simulation with that obtained from the dynamics of $\tilde{H}(t)$ in Eq.~(\ref{eq:lmg_rotated}) after truncating the series at orders $k\geq 1$. This comparison can be seen in Fig.~\ref{fig:lmg_ipr_rwa11}.
The IPR plots from the exact dynamics indicate that the first localization point, represented by the lowest  root of $J_0(\eta)$, does not show complete DMBL. However, DMBL is particularly conspicuous at large roots. The IPRs of the Floquet states obtained from the RWA dynamics exhibit large deviations from unity when away from the localization point as evidenced by the green and red curves in the middle panel of Fig.~\ref{fig:lmg_ipr_rwa11}. However, complete localization is seen in the RWA dynamics at any localization point, in contrast to the exact case in the top panel. Thus, it is necessary to incorporate  higher-order corrections into the RWA at lower localization points. The application of the first-order correction to RWA in the lower panel of Fig.~\ref{fig:lmg_ipr_rwa11} results in a curve structure that is closer to that from the exact dynamics.

\section{\label{sec:level4}Persistence of DMBL in the continuum limit}
\begin{figure}[t!]
	\centering
	\includegraphics[width = 9.0 cm]{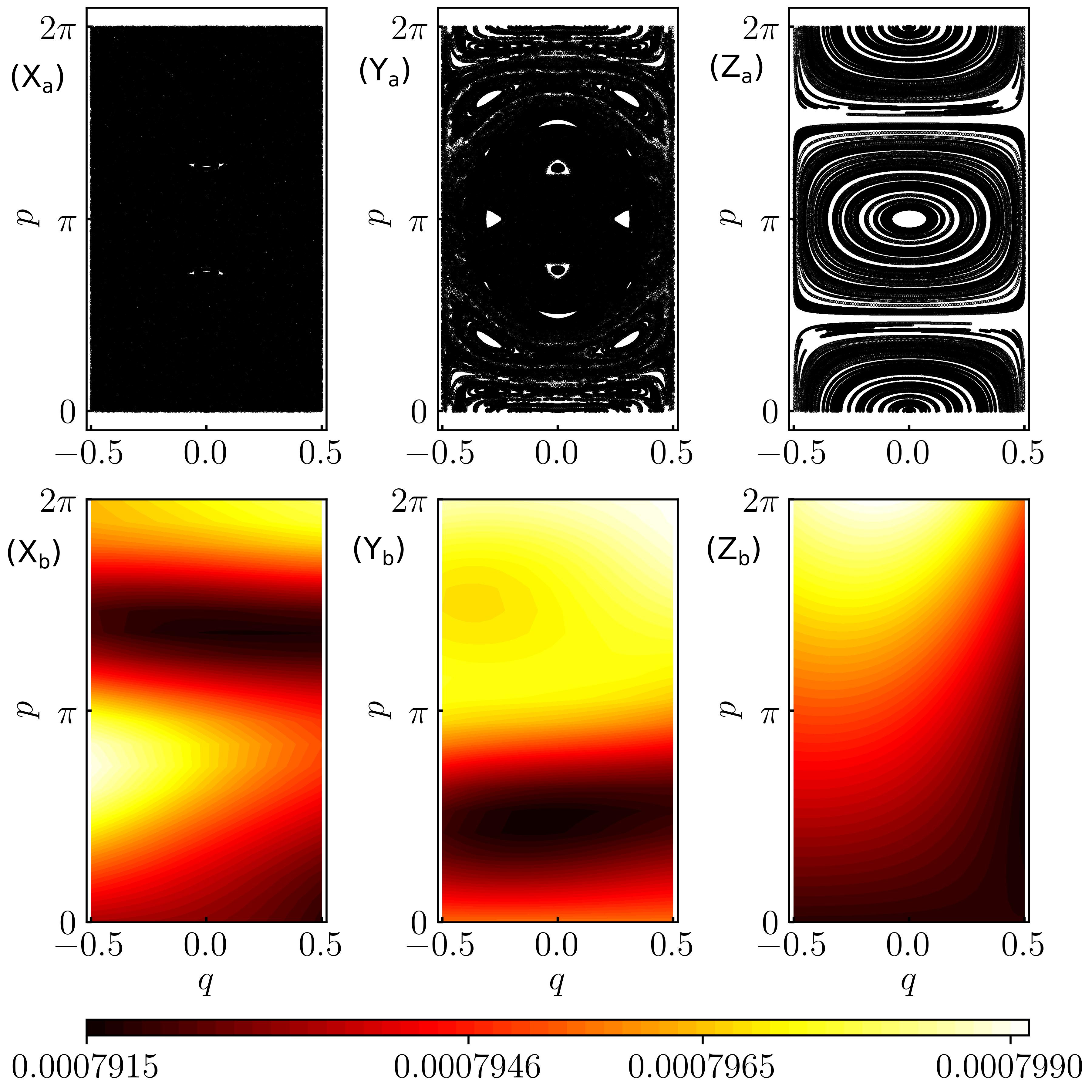}
	\caption{Phase-space distributions at $\omega=1.0$ (panels $X_{a,b}$), $\omega=2.5$ (panels $Y_{a,b}$) and $\omega=90.0$ (panels $Z_{a,b}$) for $100$ initial conditions. The drive amplitude $h$ is always adjusted such that $\eta=4h/\omega$ lies on the smallest root of $J_0(\eta)$, \textit{i.e.} $\eta=2.4048\dots$. At small $\omega=1.0$, the classical PSOS, obtained from simulating the dynamics in Eq.~(\ref{eq:class:poinc}) (panel $X_{a}$),  shows chaotic behavior, and at panel $Y_{a}$ where $\omega = 2.5$, regular regions start to appear. At higher $\omega = 90.0$, the dominance of regular dynamics can be readily seen (panel $Z_{a}$). The bottom panels plot the corresponding Spectral-Averaged Husimi-Q function, obtained from the Floquet modes $\ket{\phi^n}$ using Eq.~(\ref{eq:husimi}), and setting $N=100$. The $\omega=1.0$ case (panel $X_{b}$) has a dispersed distribution in color. This is consistent with the chaotic behavior seen in the continuum limit. At $\omega = 2.5$ (panel $Y_{b}$), there is a partial regular pattern observed together with a dispersed pattern. In the $\omega=90-$case (panel $Z_{b}$), the distribution has distinct color contrasts, which is consistent with the regular dynamics pattern seen in the continuum limit.}
	\label{fig:classical_lipkin}
\end{figure}

In the continuum limit, where $N\rightarrow\infty$, the disparity between neighboring values of $s_i$ in Eq.~(\ref{eq:lmg:tssham}) can be disregarded, and $s_i$ can be mapped to a continuum $q\in \left[-1/2, 1/2\right]$~\cite{mori_prethermalization_2019}. We define	the Hamiltonian per particle $h(t)\equiv \frac{H(t)}{N}$, and a canonically conjugate coordinate $Np\equiv \expval{-i\pdv{}{q}}$. Then, in this limit, the dynamics can be approximated by that of a classical Hamiltonian~\cite{sciolla_quantum_2010}
\begin{equation}
	h(t) = -2 q^2 - \left[h\cos{\omega t} + h_0\right]\;\sqrt{1-4q^2}\;\cos{p},
	\label{eq:class:ham}
\end{equation}
which yields the dynamical system 
\begin{align}
	\dv{q}{t} &= \pdv{h}{p} = h(t)\sqrt{1-4q^2}\sin{p}\nonumber \\
	\dv{p}{t} &= -\pdv{h}{q} = 4q\bigg[1-\frac{h(t)\cos{p}}{\sqrt{1-4q^2}}\bigg],
	\label{eq:class:poinc}
\end{align}
where $h(t) = \left[h\cos{\omega t} + h_0\right]$. We have profiled simulations of the ensuing dynamics with the \emph{Poincar$\acute{e}$ surface of section} (PSOS) of the full dynamics. Here, the $\left(q,p\right)-$phase space is strobed at $t=nT$, and plotted for many initial conditions. The results are shown in the upper panels of Fig.~\ref{fig:classical_lipkin} for a small value of $\omega=2.0$ (left panel) and a large value $\omega=90$ (right panel). In both cases, the value of $h$ is chosen such that $\eta$ lies on the first root of $J_0(\eta)$. The onset of chaos for small drive frequency indicates thermal behavior for typical initial conditions, with small islands of regularity for others. This is consistent with similar results for small frequencies reported in Refs.~[\cite{russomanno_thermalization_2015, Kidd2019}]. \emph{However,} at high frequency, the regular islands distinctly dominate over the chaos. The trajectories indicate that the conservation of $\expval{S^x}\approx \sqrt{1-4q^2}\;\cos{p}$~\cite{mori_prethermalization_2019} at high $\omega$ persists in the thermodynamic limit.
That this is a signature of the underlying quantum dynamics can be readily seen in the quantum phase space representation of the Floquet Eigenstates for a large but finite $N$. These are shown in the corresponding lower panels of Fig.~\ref{fig:classical_lipkin}. Here, we have plotted the spectral average of the Husimi Q-functions of the acquired Floquet States in the TSS. Specifically, for a coherent state $\ket{q,p}$, the corresponding spectral-averaged Husimi distribution~\cite{husimi} is obtained by 
\begin{equation}
	H(q,p)\equiv \frac{1}{\big(2N+1\big)\pi}\sum_n \ip{q,p}{\phi^n}\ip{\phi^n}{q,p}.
	\label{eq:husimi}
\end{equation}
The quantum phase space retains signatures of the classical phase space dynamics when $N=100$, indicating the onset of the persistence of $S^x$ conservation that arises from the freezing condition at high frequencies. 
\section{\label{sec:level5}Phase Crossover from Thermal to DMBL}
The analysis of the periodically driven LMG model reveals two distinct scenarios at low and high external drive frequencies. In the former case, thermalization in accordance with FETH is seen, whereas in the latter case, DMBL is induced. As a result, we hypothesize that a macroscopic change in phase occurs due to the influence of frequency. 

\begin{figure}[t!]
	\centering
	\includegraphics[width =8.5cm]{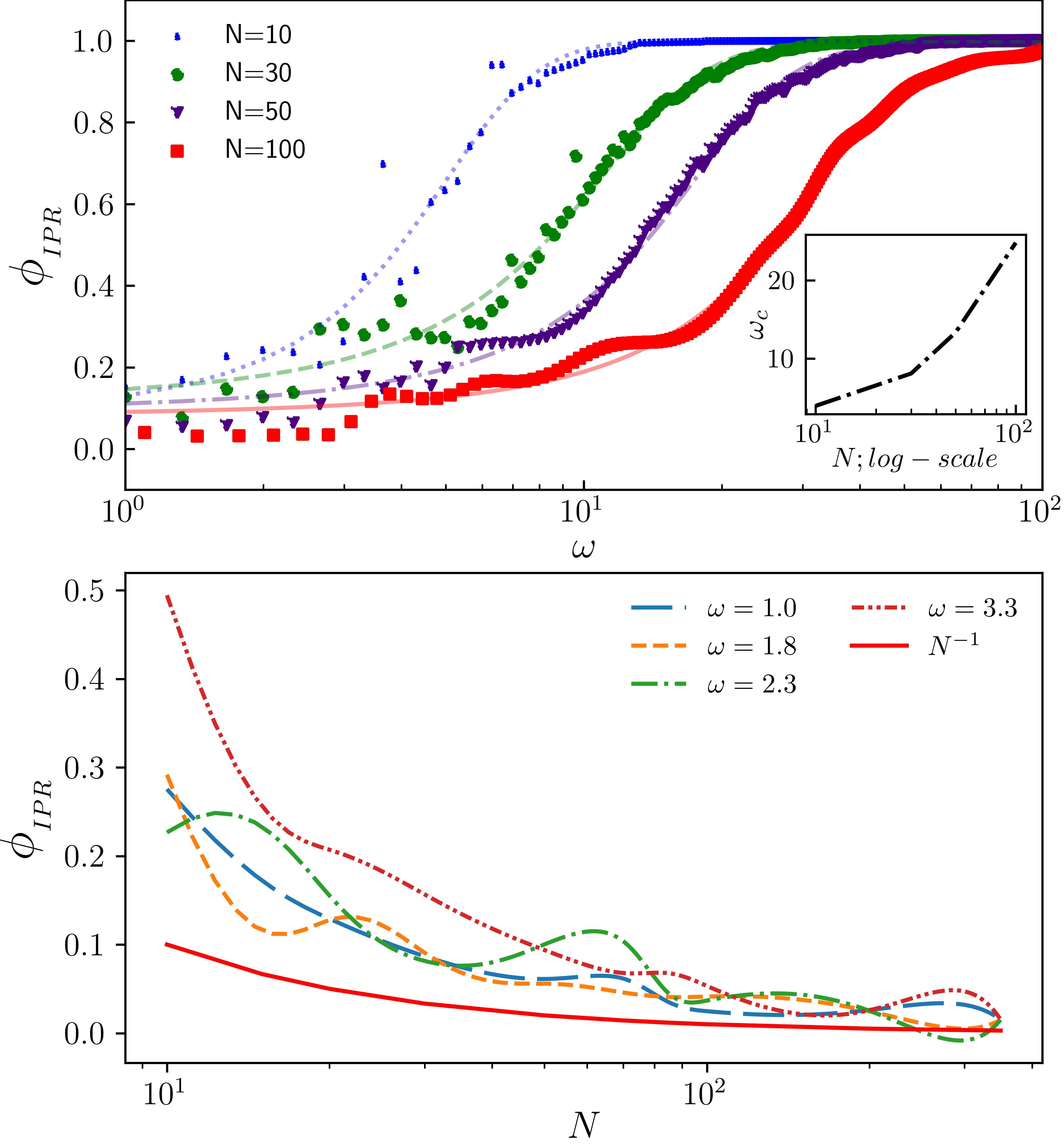}
	\caption{In the top panel, the IPR of the Floquet state with the lowest quasienergy is plotted as a function of $\omega$ for different $N$s, with amplitude $h$ adjusted to ensure that $J_0(\eta)=0$ always.The data points are shown, together with sigmoid curves $\phi_\mathrm{IPR}=\left[1+e^{-\alpha\left(\omega-\omega_c\right)}\right]^{-1}$ that were fitted to each dataset. The smooth rise in IPR defines a phase crossover (top panel) between a fully thermal phase to a fully localized phase. In the inset, the crossover frequency $\omega_c$ is plotted against $N$, and  is found to rise sharply with $N$. The bottom panel plots the IPR versus $N$ for small $\omega$s, also with $h$ adjusted. The curves asymptotically approach the dependency $\phi_\mathrm{IPR}\sim 1/N$, indicating thermalization at low-$\omega$.}
	\label{fig:phase_crossover}
\end{figure}

To demonstrate this, we investigate the IPR of the  Floquet mode with smallest quasienergy for numerous frequencies and system sizes, along with the associated drive amplitude $h$ keeping the system at a localization point. The results are shown in Fig.~\ref{fig:phase_crossover}. In the low-frequency range $\omega \in \left[1.0, 9.0\right]$, the IPR exhibits values well below unity.   {Moreover, as can be seen in the bottom panel of the same figure, when the dynamics is simulated for smaller} $\omega$s, the fall of $\phi_\mathrm{IPR}(N)$ {asymptotically approaches one that characterizes a fully thermal state, where} $\phi_\mathrm{IPR}(N)\sim 1/N$ in the TSS.  In the limit  $N\rightarrow\infty$, the IPR tends towards zero, indicating a fully delocalized state. Contrast this with the IPR plots shown in the bottom left panel of Fig.~\ref{fig:ipr:isinglowfrk} for the integrable TFIM.  The top panel of Fig.~\ref{fig:phase_crossover} also reveals a gradual increase in the unity towards unity of IPR over a certain frequency range, specifically at $\omega \approx 5$. In addition, the rise does not cross with those for different values of $N$, suggesting the onset of a phase crossover~[\cite{sierant_2023, sachdev_quantum_2011}]. As the size of the system increases, the crossover region becomes smoother, rather than sharper.  The crossover frequency $\omega_c$ have been estimated from the IPR data and plotted for different system sizes $N$ in the inset of Fig.~\ref{fig:phase_crossover}. It is observed that $\omega_c$ rises sharply as $N$ increases past $40$. This indicates that $\omega_c$ blows up in the thermodynamic limit, where $N \to \infty$.
\begin{figure}[ht!]
	\centering
	\includegraphics[width = 8.cm]{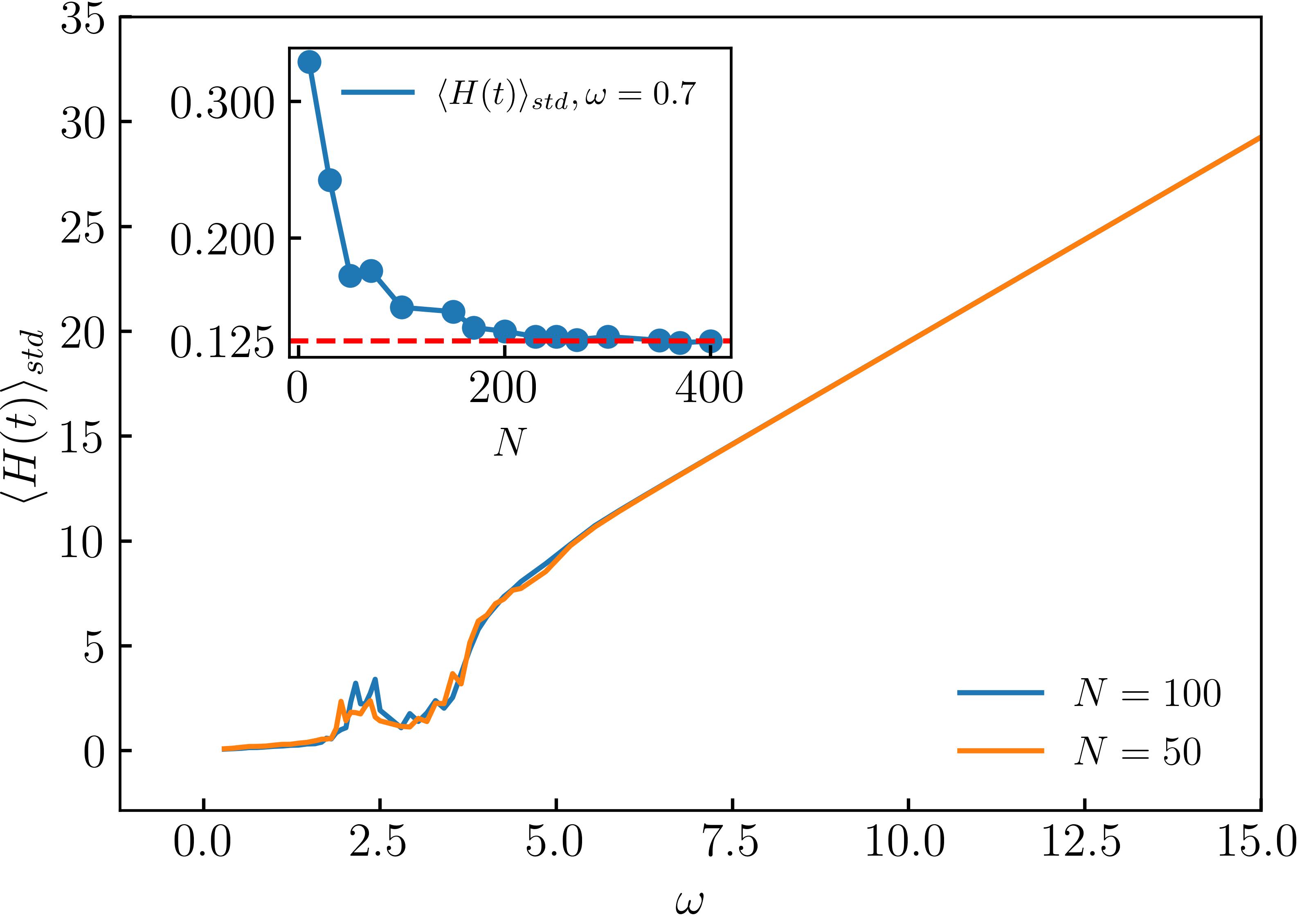}
	\caption{The temporal standard deviation of the heating rate, denoted by $\expval{H}_{std}$, calculated over a span of $t=500 T$ for two system sizes. Here, $h$ is varied to keep $\eta = 4h/\omega$ at the first root of $J_0(\eta)$. In the inset, $\expval{H}_{std}$ is plotted against system size at low-frequency ($\omega=0.7$).}
	\label{fig:havg_std}
\end{figure}

We can also look at this crossover more clearly in the plots of the heating rate of the system, defined simply by the expectation value of the Hamiltonian, $\expval{\hat{H}(t)}$. We have carried out the numerical evaluation from the simulated dynamics over $t=500 T$. When the system is adequately described by FETH, the temporal fluctuations in the {Hamiltonian}, defined by $\expval{H}^2_{std}\equiv \overline{\expval{\hat{H}}^2}-\overline{\expval{\hat{H}}}^2$ [see Eq.~(\ref{eq:lt_avg})], approach $1/8$ in the thermodynamic limit~\footnote{Thermodynamically, at $\beta\equiv 1/k_BT=0$, we have $\expval{H}^2_{std}=\Tr{H^2_0}/2^N$. Substituting eqn.~\ref{eq:h0h1} for $H_0$ yields $1/8$ on the RHS.}. Conversely, the onset of athermality is indicated by nonzero fluctuations in time. If we set the initial state to the fully polarized state in the TSS (given by $\ket{s_N}$), then the 
onset of freezing, together with DMBL, will result in nearly infinite hysteresis in the ensuing dynamics, causing $\ket{\psi}(t)\approx \ket{s_N} \forall t$. From Eq.~(\ref{eq:driven:ham}), we can clearly see that this will lead to a linearly rising dependence on $\omega$ in $\expval{H}_{std}$ as long as we stick to a localization point given by a fixed $4h/\omega$~\footnote{When frozen, $\ket{\psi(t)}\approx \ket{s_N}$. From Eq.~(\ref{eq:driven:ham}), $\expval{\hat{H}_{0,1}}$ are both approx. constant. Averaging the square of $\expval{\hat{H}(t)}$ over long times yields dependency  $\sim h^2 + \delta$, where $\delta\sim h_0\ll 1$. Thus, the std. devn. in time $\sim h\sim \omega$, since $\eta=4h/\omega$ is kept fixed.}. All these observations are corroborated by the plots in Fig.~\ref{fig:havg_std}, where we have displayed the temporal standard deviation of the heating rate for increasing $\omega$, while engineering $h$ such that the system is always at a localization point. A nonsingular rise has been identified at $\omega \approx 4.0$. The heating rate exhibits diminutive fluctuations below that value of $\omega$, consistent with thermalization, while a linear rise is observed at higher frequencies, consistent with actual localization. When $\omega \ll 4.0$, the diminutive standard deviation asymptotically approaches the theoretical value of $0.125$ in the thermodynamic limit, as can be seen in the inset of Fig.~\ref{fig:havg_std}. The small peaks observed at $\omega\in [2, 4)$ are finite-size effects that disappear in the thermodynamic limit. In the inset, $\expval{H}_{std}$ is plotted against system size at low-frequency ($\omega=0.7$) and $h$ engineered to lie at a localization point.

\section{\label{sec:level7}Conclusion and Outlook}
We have delved into the onset of freezing and phase cross-over in 1D spin systems driven by a time periodic transverse field, contrasting the responses in the TFIM with that of the LMG. The parametrization of DMBL is based on the IPR of the Floquet eigenstates. Our investigations compared the IPRs from both models numerically, and found the emergence of thermal behavior at low frequencies and freezing at high frequencies for the LMG model, the latter being a direct consequence of the appearance of additional approximately conserved quantities.

Long-range spins exhibit strong localization in spin-coordinate space for the LMG model when the drive frequency is $\omega \gg J$, where $J$ represents the spin exchange energy. The localization of the LMG model occurs at specific freezing points in the space of the drive frequency $\omega$ and amplitude $h$, at $J_0(4h/\omega)=0,\; \omega \gg J$. This is apparently similar to the phenomenon of DMF in the TFIM, as well as DMBL in non-integrable short-range models. Plots of the IPR for a range of frequencies along the freezing manifold exhibits a smooth increase in IPR yielding a quantum phase-crossover from a thermal phase governed by the FETH to a DMBL phase. However, in sharp contrast to the phase transitions seen in short-range models, the crossover frequency $\omega_c$ diverges with the system size, indicating that DMBL is unstable in the thermodynamic limit for finite drive frequencies. Nonetheless, the suppression  of thermalization through  dynamical many-body localization in finite sized long-range systems can be controlled via Floquet engineering, even in clean systems without any disorder.	Thus, periodically driven long-range spin systems are an excellent tool for investigating disorder-free many-body localization, as can be readily seen via the IPR of its Floquet modes.

There are several unexplored indicators of DMBL, such as entanglement entropy and level statistics~\cite{khemani_phase_2016}, which we defer to future studies. In addition, Halpern in 2019 proposed a quantum engine based on MBL\cite{yunger_halpern_quantum_2019} which works  between strong localized and thermal phases of the system. In our proposed LMG model, tuning the system parameters by bringing them to the freezing points, then adiabatically cycling the frequency from the thermal region to the DMBL region, can achieve a similar engine without going through a phase transition. 

\begin{acknowledgments}
MR acknowledges The University of Burdwan for support via a state-funded fellowship. AR acknowledges support from the University Grants Commission (UGC) of India, Grant No. F.30-425/2018(BSR), as well as from the Science and Engineering Research Board (SERB), Grant No. CRG/20l8/004002.
AR thanks Prof. Arnab Das, School of Physical Sciences, Indian Association for the Cultivation of Science, Kolkata, for useful suggestions.
\end{acknowledgments}

\bibliographystyle{apsrev4-2}
\bibliography{main}

\begin{thebibliography}{72}%
\makeatletter
\providecommand \@ifxundefined [1]{%
 \@ifx{#1\undefined}
}%
\providecommand \@ifnum [1]{%
 \ifnum #1\expandafter \@firstoftwo
 \else \expandafter \@secondoftwo
 \fi
}%
\providecommand \@ifx [1]{%
 \ifx #1\expandafter \@firstoftwo
 \else \expandafter \@secondoftwo
 \fi
}%
\providecommand \natexlab [1]{#1}%
\providecommand \enquote  [1]{``#1''}%
\providecommand \bibnamefont  [1]{#1}%
\providecommand \bibfnamefont [1]{#1}%
\providecommand \citenamefont [1]{#1}%
\providecommand \href@noop [0]{\@secondoftwo}%
\providecommand \href [0]{\begingroup \@sanitize@url \@href}%
\providecommand \@href[1]{\@@startlink{#1}\@@href}%
\providecommand \@@href[1]{\endgroup#1\@@endlink}%
\providecommand \@sanitize@url [0]{\catcode `\\12\catcode `\$12\catcode
  `\&12\catcode `\#12\catcode `\^12\catcode `\_12\catcode `\%12\relax}%
\providecommand \@@startlink[1]{}%
\providecommand \@@endlink[0]{}%
\providecommand \url  [0]{\begingroup\@sanitize@url \@url }%
\providecommand \@url [1]{\endgroup\@href {#1}{\urlprefix }}%
\providecommand \urlprefix  [0]{URL }%
\providecommand \Eprint [0]{\href }%
\providecommand \doibase [0]{https://doi.org/}%
\providecommand \selectlanguage [0]{\@gobble}%
\providecommand \bibinfo  [0]{\@secondoftwo}%
\providecommand \bibfield  [0]{\@secondoftwo}%
\providecommand \translation [1]{[#1]}%
\providecommand \BibitemOpen [0]{}%
\providecommand \bibitemStop [0]{}%
\providecommand \bibitemNoStop [0]{.\EOS\space}%
\providecommand \EOS [0]{\spacefactor3000\relax}%
\providecommand \BibitemShut  [1]{\csname bibitem#1\endcsname}%
\let\auto@bib@innerbib\@empty
\bibitem [{\citenamefont {Bordia}\ \emph {et~al.}(2017)\citenamefont {Bordia},
  \citenamefont {L{\"u}schen}, \citenamefont {Schneider}, \citenamefont
  {Knap},\ and\ \citenamefont {Bloch}}]{bordia_periodically_2017}%
  \BibitemOpen
  \bibfield  {author} {\bibinfo {author} {\bibfnamefont {P.}~\bibnamefont
  {Bordia}}, \bibinfo {author} {\bibfnamefont {H.}~\bibnamefont {L{\"u}schen}},
  \bibinfo {author} {\bibfnamefont {U.}~\bibnamefont {Schneider}}, \bibinfo
  {author} {\bibfnamefont {M.}~\bibnamefont {Knap}},\ and\ \bibinfo {author}
  {\bibfnamefont {I.}~\bibnamefont {Bloch}},\ }\href
  {https://doi.org/10.1038/nphys4020} {\bibfield  {journal} {\bibinfo
  {journal} {Nat. Phys.}\ }\textbf {\bibinfo {volume} {13}},\ \bibinfo {pages}
  {460} (\bibinfo {year} {2017})}\BibitemShut {NoStop}%
\bibitem [{\citenamefont {Sahoo}\ \emph {et~al.}(2019)\citenamefont {Sahoo},
  \citenamefont {Schneider},\ and\ \citenamefont
  {Eggert}}]{sahoo_periodically_2019}%
  \BibitemOpen
  \bibfield  {author} {\bibinfo {author} {\bibfnamefont {S.}~\bibnamefont
  {Sahoo}}, \bibinfo {author} {\bibfnamefont {I.}~\bibnamefont {Schneider}},\
  and\ \bibinfo {author} {\bibfnamefont {S.}~\bibnamefont {Eggert}},\
  }\href@noop {} {} (\bibinfo {year} {2019}),\ \Eprint
  {https://arxiv.org/abs/1906.00004} {arXiv:1906.00004 [cond-mat.str-el]}
  \BibitemShut {NoStop}%
\bibitem [{\citenamefont {Das}(2010)}]{das_exotic_2010}%
  \BibitemOpen
  \bibfield  {author} {\bibinfo {author} {\bibfnamefont {A.}~\bibnamefont
  {Das}},\ }\href {https://doi.org/10.1103/PhysRevB.82.172402} {\bibfield
  {journal} {\bibinfo  {journal} {Phys. Rev. B}\ }\textbf {\bibinfo {volume}
  {82}},\ \bibinfo {pages} {172402} (\bibinfo {year} {2010})}\BibitemShut
  {NoStop}%
\bibitem [{\citenamefont {Bhattacharyya}\ \emph {et~al.}(2012)\citenamefont
  {Bhattacharyya}, \citenamefont {Das},\ and\ \citenamefont
  {Dasgupta}}]{sirshendu:freezing}%
  \BibitemOpen
  \bibfield  {author} {\bibinfo {author} {\bibfnamefont {S.}~\bibnamefont
  {Bhattacharyya}}, \bibinfo {author} {\bibfnamefont {A.}~\bibnamefont {Das}},\
  and\ \bibinfo {author} {\bibfnamefont {S.}~\bibnamefont {Dasgupta}},\ }\href
  {https://doi.org/10.1103/PhysRevB.86.054410} {\bibfield  {journal} {\bibinfo
  {journal} {Phys. Rev. B}\ }\textbf {\bibinfo {volume} {86}},\ \bibinfo
  {pages} {054410} (\bibinfo {year} {2012})}\BibitemShut {NoStop}%
\bibitem [{\citenamefont {Haldar}\ \emph {et~al.}(2018)\citenamefont {Haldar},
  \citenamefont {Moessner},\ and\ \citenamefont
  {Das}}]{asmi:floquet:thermalization}%
  \BibitemOpen
  \bibfield  {author} {\bibinfo {author} {\bibfnamefont {A.}~\bibnamefont
  {Haldar}}, \bibinfo {author} {\bibfnamefont {R.}~\bibnamefont {Moessner}},\
  and\ \bibinfo {author} {\bibfnamefont {A.}~\bibnamefont {Das}},\ }\href
  {https://doi.org/10.1103/PhysRevB.97.245122} {\bibfield  {journal} {\bibinfo
  {journal} {Phys. Rev. B}\ }\textbf {\bibinfo {volume} {97}},\ \bibinfo
  {pages} {245122} (\bibinfo {year} {2018})}\BibitemShut {NoStop}%
\bibitem [{\citenamefont {Haldar}\ \emph {et~al.}(2021)\citenamefont {Haldar},
  \citenamefont {Sen}, \citenamefont {Moessner},\ and\ \citenamefont
  {Das}}]{asmi:scars}%
  \BibitemOpen
  \bibfield  {author} {\bibinfo {author} {\bibfnamefont {A.}~\bibnamefont
  {Haldar}}, \bibinfo {author} {\bibfnamefont {D.}~\bibnamefont {Sen}},
  \bibinfo {author} {\bibfnamefont {R.}~\bibnamefont {Moessner}},\ and\
  \bibinfo {author} {\bibfnamefont {A.}~\bibnamefont {Das}},\ }\href
  {https://doi.org/10.1103/PhysRevX.11.021008} {\bibfield  {journal} {\bibinfo
  {journal} {Phys. Rev. X}\ }\textbf {\bibinfo {volume} {11}},\ \bibinfo
  {pages} {021008} (\bibinfo {year} {2021})}\BibitemShut {NoStop}%
\bibitem [{\citenamefont {Mbeng}\ \emph {et~al.}(2020)\citenamefont {Mbeng},
  \citenamefont {Russomanno},\ and\ \citenamefont
  {Santoro}}]{mbeng_quantum_2020}%
  \BibitemOpen
  \bibfield  {author} {\bibinfo {author} {\bibfnamefont {G.~B.}\ \bibnamefont
  {Mbeng}}, \bibinfo {author} {\bibfnamefont {A.}~\bibnamefont {Russomanno}},\
  and\ \bibinfo {author} {\bibfnamefont {G.~E.}\ \bibnamefont {Santoro}},\
  }\href@noop {} {} (\bibinfo {year} {2020}),\ \Eprint
  {https://arxiv.org/abs/2009.09208} {arXiv:2009.09208 [quant-ph]} \BibitemShut
  {NoStop}%
\bibitem [{\citenamefont {Yamada}\ and\ \citenamefont
  {Ikeda}(2022)}]{yamada_localization_2022}%
  \BibitemOpen
  \bibfield  {author} {\bibinfo {author} {\bibfnamefont {H.~S.}\ \bibnamefont
  {Yamada}}\ and\ \bibinfo {author} {\bibfnamefont {K.~S.}\ \bibnamefont
  {Ikeda}},\ }\href {https://doi.org/10.1103/PhysRevE.105.054201} {\bibfield
  {journal} {\bibinfo  {journal} {Phys. Rev. E}\ }\textbf {\bibinfo {volume}
  {105}},\ \bibinfo {pages} {054201} (\bibinfo {year} {2022})}\BibitemShut
  {NoStop}%
\bibitem [{\citenamefont {Roy}\ and\ \citenamefont
  {Das}(2015)}]{roy_fate_2015}%
  \BibitemOpen
  \bibfield  {author} {\bibinfo {author} {\bibfnamefont {A.}~\bibnamefont
  {Roy}}\ and\ \bibinfo {author} {\bibfnamefont {A.}~\bibnamefont {Das}},\
  }\href {https://doi.org/10.1103/PhysRevB.91.121106} {\bibfield  {journal}
  {\bibinfo  {journal} {Phys. Rev. B}\ }\textbf {\bibinfo {volume} {91}},\
  \bibinfo {pages} {121106(R)} (\bibinfo {year} {2015})}\BibitemShut {NoStop}%
\bibitem [{\citenamefont {Li}\ \emph {et~al.}(2018)\citenamefont {Li},
  \citenamefont {Shapiro},\ and\ \citenamefont {Kottos}}]{li_floquet_2018}%
  \BibitemOpen
  \bibfield  {author} {\bibinfo {author} {\bibfnamefont {H.}~\bibnamefont
  {Li}}, \bibinfo {author} {\bibfnamefont {B.}~\bibnamefont {Shapiro}},\ and\
  \bibinfo {author} {\bibfnamefont {T.}~\bibnamefont {Kottos}},\ }\href
  {https://doi.org/10.1103/PhysRevB.98.121101} {\bibfield  {journal} {\bibinfo
  {journal} {Phys. Rev. B}\ }\textbf {\bibinfo {volume} {98}},\ \bibinfo
  {pages} {121101(R)} (\bibinfo {year} {2018})}\BibitemShut {NoStop}%
\bibitem [{\citenamefont {Eckardt}\ and\ \citenamefont
  {Anisimovas}(2015)}]{eckardt_high_frequency_2015}%
  \BibitemOpen
  \bibfield  {author} {\bibinfo {author} {\bibfnamefont {A.}~\bibnamefont
  {Eckardt}}\ and\ \bibinfo {author} {\bibfnamefont {E.}~\bibnamefont
  {Anisimovas}},\ }\href {https://doi.org/10.1088/1367-2630/17/9/093039}
  {\bibfield  {journal} {\bibinfo  {journal} {New J. Phys.}\ }\textbf {\bibinfo
  {volume} {17}},\ \bibinfo {pages} {093039} (\bibinfo {year}
  {2015})}\BibitemShut {NoStop}%
\bibitem [{\citenamefont {Zhang}\ \emph {et~al.}(2016)\citenamefont {Zhang},
  \citenamefont {Khemani},\ and\ \citenamefont {Huse}}]{zhang_floquet_2016}%
  \BibitemOpen
  \bibfield  {author} {\bibinfo {author} {\bibfnamefont {L.}~\bibnamefont
  {Zhang}}, \bibinfo {author} {\bibfnamefont {V.}~\bibnamefont {Khemani}},\
  and\ \bibinfo {author} {\bibfnamefont {D.~A.}\ \bibnamefont {Huse}},\ }\href
  {https://doi.org/10.1103/PhysRevB.94.224202} {\bibfield  {journal} {\bibinfo
  {journal} {Phys. Rev. B}\ }\textbf {\bibinfo {volume} {94}},\ \bibinfo
  {pages} {224202} (\bibinfo {year} {2016})}\BibitemShut {NoStop}%
\bibitem [{\citenamefont {Mori}\ \emph {et~al.}(2018)\citenamefont {Mori},
  \citenamefont {Ikeda}, \citenamefont {Kaminishi},\ and\ \citenamefont
  {Ueda}}]{Mori_2018}%
  \BibitemOpen
  \bibfield  {author} {\bibinfo {author} {\bibfnamefont {T.}~\bibnamefont
  {Mori}}, \bibinfo {author} {\bibfnamefont {T.~N.}\ \bibnamefont {Ikeda}},
  \bibinfo {author} {\bibfnamefont {E.}~\bibnamefont {Kaminishi}},\ and\
  \bibinfo {author} {\bibfnamefont {M.}~\bibnamefont {Ueda}},\ }\href
  {https://doi.org/10.1088/1361-6455/aabcdf} {\bibfield  {journal} {\bibinfo
  {journal} {Journal of Physics B: Atomic, Molecular and Optical Physics}\
  }\textbf {\bibinfo {volume} {51}},\ \bibinfo {pages} {112001} (\bibinfo
  {year} {2018})}\BibitemShut {NoStop}%
\bibitem [{\citenamefont {Kim}\ \emph {et~al.}(2014)\citenamefont {Kim},
  \citenamefont {Ikeda},\ and\ \citenamefont {Huse}}]{Kim_2014}%
  \BibitemOpen
  \bibfield  {author} {\bibinfo {author} {\bibfnamefont {H.}~\bibnamefont
  {Kim}}, \bibinfo {author} {\bibfnamefont {T.~N.}\ \bibnamefont {Ikeda}},\
  and\ \bibinfo {author} {\bibfnamefont {D.~A.}\ \bibnamefont {Huse}},\ }\href
  {https://doi.org/10.1103/PhysRevE.90.052105} {\bibfield  {journal} {\bibinfo
  {journal} {Phys. Rev. E}\ }\textbf {\bibinfo {volume} {90}},\ \bibinfo
  {pages} {052105} (\bibinfo {year} {2014})}\BibitemShut {NoStop}%
\bibitem [{\citenamefont {Mizuta}\ \emph {et~al.}(2020)\citenamefont {Mizuta},
  \citenamefont {Takasan},\ and\ \citenamefont {Kawakami}}]{Mizuta_2020}%
  \BibitemOpen
  \bibfield  {author} {\bibinfo {author} {\bibfnamefont {K.}~\bibnamefont
  {Mizuta}}, \bibinfo {author} {\bibfnamefont {K.}~\bibnamefont {Takasan}},\
  and\ \bibinfo {author} {\bibfnamefont {N.}~\bibnamefont {Kawakami}},\ }\href
  {https://doi.org/10.1103/PhysRevResearch.2.033284} {\bibfield  {journal}
  {\bibinfo  {journal} {Phys. Rev. Res.}\ }\textbf {\bibinfo {volume} {2}},\
  \bibinfo {pages} {033284} (\bibinfo {year} {2020})}\BibitemShut {NoStop}%
\bibitem [{\citenamefont {Mori}(2023)}]{Mori_2023_1}%
  \BibitemOpen
  \bibfield  {author} {\bibinfo {author} {\bibfnamefont {T.}~\bibnamefont
  {Mori}},\ }\href {https://doi.org/10.1146/annurev-conmatphys-040721-015537}
  {\bibfield  {journal} {\bibinfo  {journal} {Annual Review of Condensed Matter
  Physics}\ }\textbf {\bibinfo {volume} {14}},\ \bibinfo {pages} {35} (\bibinfo
  {year} {2023})}\BibitemShut {NoStop}%
\bibitem [{\citenamefont {Khemani}\ \emph {et~al.}(2016)\citenamefont
  {Khemani}, \citenamefont {Lazarides}, \citenamefont {Moessner},\ and\
  \citenamefont {Sondhi}}]{khemani_phase_2016}%
  \BibitemOpen
  \bibfield  {author} {\bibinfo {author} {\bibfnamefont {V.}~\bibnamefont
  {Khemani}}, \bibinfo {author} {\bibfnamefont {A.}~\bibnamefont {Lazarides}},
  \bibinfo {author} {\bibfnamefont {R.}~\bibnamefont {Moessner}},\ and\
  \bibinfo {author} {\bibfnamefont {S.~L.}\ \bibnamefont {Sondhi}},\ }\href
  {https://doi.org/10.1103/PhysRevLett.116.250401} {\bibfield  {journal}
  {\bibinfo  {journal} {Phys. Rev. Lett.}\ }\textbf {\bibinfo {volume} {116}},\
  \bibinfo {pages} {250401} (\bibinfo {year} {2016})}\BibitemShut {NoStop}%
\bibitem [{\citenamefont {Yunger~Halpern}\ \emph {et~al.}(2019)\citenamefont
  {Yunger~Halpern}, \citenamefont {White}, \citenamefont {Gopalakrishnan},\
  and\ \citenamefont {Refael}}]{yunger_halpern_quantum_2019}%
  \BibitemOpen
  \bibfield  {author} {\bibinfo {author} {\bibfnamefont {N.}~\bibnamefont
  {Yunger~Halpern}}, \bibinfo {author} {\bibfnamefont {C.~D.}\ \bibnamefont
  {White}}, \bibinfo {author} {\bibfnamefont {S.}~\bibnamefont
  {Gopalakrishnan}},\ and\ \bibinfo {author} {\bibfnamefont {G.}~\bibnamefont
  {Refael}},\ }\href {https://doi.org/10.1103/PhysRevB.99.024203} {\bibfield
  {journal} {\bibinfo  {journal} {Phys. Rev. B}\ }\textbf {\bibinfo {volume}
  {99}},\ \bibinfo {pages} {024203} (\bibinfo {year} {2019})}\BibitemShut
  {NoStop}%
\bibitem [{\citenamefont {Nag}\ \emph {et~al.}(2014)\citenamefont {Nag},
  \citenamefont {Roy}, \citenamefont {Dutta},\ and\ \citenamefont
  {Sen}}]{diptiman2014}%
  \BibitemOpen
  \bibfield  {author} {\bibinfo {author} {\bibfnamefont {T.}~\bibnamefont
  {Nag}}, \bibinfo {author} {\bibfnamefont {S.}~\bibnamefont {Roy}}, \bibinfo
  {author} {\bibfnamefont {A.}~\bibnamefont {Dutta}},\ and\ \bibinfo {author}
  {\bibfnamefont {D.}~\bibnamefont {Sen}},\ }\href
  {https://doi.org/10.1103/PhysRevB.89.165425} {\bibfield  {journal} {\bibinfo
  {journal} {Phys. Rev. B}\ }\textbf {\bibinfo {volume} {89}},\ \bibinfo
  {pages} {165425} (\bibinfo {year} {2014})}\BibitemShut {NoStop}%
\bibitem [{\citenamefont {Carleo}\ \emph {et~al.}(2012)\citenamefont {Carleo},
  \citenamefont {Becca}, \citenamefont {Schir{\'o}},\ and\ \citenamefont
  {Fabrizio}}]{Carleo2012}%
  \BibitemOpen
  \bibfield  {author} {\bibinfo {author} {\bibfnamefont {G.}~\bibnamefont
  {Carleo}}, \bibinfo {author} {\bibfnamefont {F.}~\bibnamefont {Becca}},
  \bibinfo {author} {\bibfnamefont {M.}~\bibnamefont {Schir{\'o}}},\ and\
  \bibinfo {author} {\bibfnamefont {M.}~\bibnamefont {Fabrizio}},\ }\href
  {https://doi.org/10.1038/srep00243} {\bibfield  {journal} {\bibinfo
  {journal} {Sci. Rep.}\ }\textbf {\bibinfo {volume} {2}},\ \bibinfo {pages}
  {243} (\bibinfo {year} {2012})}\BibitemShut {NoStop}%
\bibitem [{\citenamefont {Aditya}\ and\ \citenamefont
  {Sen}(2023)}]{aditya2023dynamical}%
  \BibitemOpen
  \bibfield  {author} {\bibinfo {author} {\bibfnamefont {S.}~\bibnamefont
  {Aditya}}\ and\ \bibinfo {author} {\bibfnamefont {D.}~\bibnamefont {Sen}},\
  }\href {https://doi.org/10.21468/SciPostPhysCore.6.4.083} {\bibfield
  {journal} {\bibinfo  {journal} {SciPost Phys. Core}\ }\textbf {\bibinfo
  {volume} {6}},\ \bibinfo {pages} {083} (\bibinfo {year} {2023})}\BibitemShut
  {NoStop}%
\bibitem [{\citenamefont {Schiulaz}\ \emph {et~al.}(2015)\citenamefont
  {Schiulaz}, \citenamefont {Silva},\ and\ \citenamefont
  {M\"uller}}]{alessandro_markus}%
  \BibitemOpen
  \bibfield  {author} {\bibinfo {author} {\bibfnamefont {M.}~\bibnamefont
  {Schiulaz}}, \bibinfo {author} {\bibfnamefont {A.}~\bibnamefont {Silva}},\
  and\ \bibinfo {author} {\bibfnamefont {M.}~\bibnamefont {M\"uller}},\ }\href
  {https://doi.org/10.1103/PhysRevB.91.184202} {\bibfield  {journal} {\bibinfo
  {journal} {Phys. Rev. B}\ }\textbf {\bibinfo {volume} {91}},\ \bibinfo
  {pages} {184202} (\bibinfo {year} {2015})}\BibitemShut {NoStop}%
\bibitem [{\citenamefont {Grover}\ and\ \citenamefont
  {Fisher}(2014)}]{Grover2014}%
  \BibitemOpen
  \bibfield  {author} {\bibinfo {author} {\bibfnamefont {T.}~\bibnamefont
  {Grover}}\ and\ \bibinfo {author} {\bibfnamefont {M.~P.~A.}\ \bibnamefont
  {Fisher}},\ }\href {https://doi.org/10.1088/1742-5468/2014/10/P10010}
  {\bibfield  {journal} {\bibinfo  {journal} {J. Stat. Mech. Theory Exp.}\
  }\textbf {\bibinfo {volume} {2014}},\ \bibinfo {pages} {P10010} (\bibinfo
  {year} {2014})}\BibitemShut {NoStop}%
\bibitem [{\citenamefont {Papić}\ \emph {et~al.}(2015)\citenamefont {Papić},
  \citenamefont {Stoudenmire},\ and\ \citenamefont {Abanin}}]{miles2015}%
  \BibitemOpen
  \bibfield  {author} {\bibinfo {author} {\bibfnamefont {Z.}~\bibnamefont
  {Papić}}, \bibinfo {author} {\bibfnamefont {E.~M.}\ \bibnamefont
  {Stoudenmire}},\ and\ \bibinfo {author} {\bibfnamefont {D.~A.}\ \bibnamefont
  {Abanin}},\ }\href
  {https://doi.org/https://doi.org/10.1016/j.aop.2015.08.024} {\bibfield
  {journal} {\bibinfo  {journal} {Ann. Physics}\ }\textbf {\bibinfo {volume}
  {362}},\ \bibinfo {pages} {714} (\bibinfo {year} {2015})}\BibitemShut
  {NoStop}%
\bibitem [{\citenamefont {Smith}\ \emph {et~al.}(2017)\citenamefont {Smith},
  \citenamefont {Knolle}, \citenamefont {Kovrizhin},\ and\ \citenamefont
  {Moessner}}]{Smith2017}%
  \BibitemOpen
  \bibfield  {author} {\bibinfo {author} {\bibfnamefont {A.}~\bibnamefont
  {Smith}}, \bibinfo {author} {\bibfnamefont {J.}~\bibnamefont {Knolle}},
  \bibinfo {author} {\bibfnamefont {D.~L.}\ \bibnamefont {Kovrizhin}},\ and\
  \bibinfo {author} {\bibfnamefont {R.}~\bibnamefont {Moessner}},\ }\href
  {https://doi.org/10.1103/PhysRevLett.118.266601} {\bibfield  {journal}
  {\bibinfo  {journal} {Phys. Rev. Lett.}\ }\textbf {\bibinfo {volume} {118}},\
  \bibinfo {pages} {266601} (\bibinfo {year} {2017})}\BibitemShut {NoStop}%
\bibitem [{\citenamefont {Hart}\ \emph {et~al.}(2021)\citenamefont {Hart},
  \citenamefont {Gopalakrishnan},\ and\ \citenamefont
  {Castelnovo}}]{MBL_emergent_disorder}%
  \BibitemOpen
  \bibfield  {author} {\bibinfo {author} {\bibfnamefont {O.}~\bibnamefont
  {Hart}}, \bibinfo {author} {\bibfnamefont {S.}~\bibnamefont
  {Gopalakrishnan}},\ and\ \bibinfo {author} {\bibfnamefont {C.}~\bibnamefont
  {Castelnovo}},\ }\href {https://doi.org/10.1103/PhysRevLett.126.227202}
  {\bibfield  {journal} {\bibinfo  {journal} {Phys. Rev. Lett.}\ }\textbf
  {\bibinfo {volume} {126}},\ \bibinfo {pages} {227202} (\bibinfo {year}
  {2021})}\BibitemShut {NoStop}%
\bibitem [{\citenamefont {Engelhardt}\ \emph {et~al.}(2013)\citenamefont
  {Engelhardt}, \citenamefont {Bastidas}, \citenamefont {Emary},\ and\
  \citenamefont {Brandes}}]{Engelhardt2013}%
  \BibitemOpen
  \bibfield  {author} {\bibinfo {author} {\bibfnamefont {G.}~\bibnamefont
  {Engelhardt}}, \bibinfo {author} {\bibfnamefont {V.~M.}\ \bibnamefont
  {Bastidas}}, \bibinfo {author} {\bibfnamefont {C.}~\bibnamefont {Emary}},\
  and\ \bibinfo {author} {\bibfnamefont {T.}~\bibnamefont {Brandes}},\ }\href
  {https://doi.org/10.1103/PhysRevE.87.052110} {\bibfield  {journal} {\bibinfo
  {journal} {Phys. Rev. E}\ }\textbf {\bibinfo {volume} {87}},\ \bibinfo
  {pages} {052110} (\bibinfo {year} {2013})}\BibitemShut {NoStop}%
\bibitem [{\citenamefont {Lipkin}\ \emph {et~al.}(1965)\citenamefont {Lipkin},
  \citenamefont {Meshkov},\ and\ \citenamefont {Glick}}]{lmg1965_1}%
  \BibitemOpen
  \bibfield  {author} {\bibinfo {author} {\bibfnamefont {H.}~\bibnamefont
  {Lipkin}}, \bibinfo {author} {\bibfnamefont {N.}~\bibnamefont {Meshkov}},\
  and\ \bibinfo {author} {\bibfnamefont {A.}~\bibnamefont {Glick}},\ }\href
  {https://doi.org/https://doi.org/10.1016/0029-5582(65)90862-X} {\bibfield
  {journal} {\bibinfo  {journal} {Nuclear Phys. B}\ }\textbf {\bibinfo {volume}
  {62}},\ \bibinfo {pages} {188} (\bibinfo {year} {1965})}\BibitemShut
  {NoStop}%
\bibitem [{\citenamefont {Meshkov}\ \emph {et~al.}(1965)\citenamefont
  {Meshkov}, \citenamefont {Glick},\ and\ \citenamefont {Lipkin}}]{lmg1965_2}%
  \BibitemOpen
  \bibfield  {author} {\bibinfo {author} {\bibfnamefont {N.}~\bibnamefont
  {Meshkov}}, \bibinfo {author} {\bibfnamefont {A.}~\bibnamefont {Glick}},\
  and\ \bibinfo {author} {\bibfnamefont {H.}~\bibnamefont {Lipkin}},\ }\href
  {https://doi.org/https://doi.org/10.1016/0029-5582(65)90863-1} {\bibfield
  {journal} {\bibinfo  {journal} {Nuclear Phys. B}\ }\textbf {\bibinfo {volume}
  {62}},\ \bibinfo {pages} {199} (\bibinfo {year} {1965})}\BibitemShut
  {NoStop}%
\bibitem [{\citenamefont {Glick}\ \emph {et~al.}(1965)\citenamefont {Glick},
  \citenamefont {Lipkin},\ and\ \citenamefont {Meshkov}}]{lmg1965_3}%
  \BibitemOpen
  \bibfield  {author} {\bibinfo {author} {\bibfnamefont {A.}~\bibnamefont
  {Glick}}, \bibinfo {author} {\bibfnamefont {H.}~\bibnamefont {Lipkin}},\ and\
  \bibinfo {author} {\bibfnamefont {N.}~\bibnamefont {Meshkov}},\ }\href
  {https://doi.org/https://doi.org/10.1016/0029-5582(65)90864-3} {\bibfield
  {journal} {\bibinfo  {journal} {Nuclear Phys. B}\ }\textbf {\bibinfo {volume}
  {62}},\ \bibinfo {pages} {211} (\bibinfo {year} {1965})}\BibitemShut
  {NoStop}%
\bibitem [{\citenamefont {Debergh}\ and\ \citenamefont
  {Stancu}(2001)}]{debergh_2001}%
  \BibitemOpen
  \bibfield  {author} {\bibinfo {author} {\bibfnamefont {N.}~\bibnamefont
  {Debergh}}\ and\ \bibinfo {author} {\bibfnamefont {F.}~\bibnamefont
  {Stancu}},\ }\href {https://doi.org/10.1088/0305-4470/34/15/305} {\bibfield
  {journal} {\bibinfo  {journal} {J. Phys. A: Math. Gen.}\ }\textbf {\bibinfo
  {volume} {34}},\ \bibinfo {pages} {3265} (\bibinfo {year}
  {2001})}\BibitemShut {NoStop}%
\bibitem [{\citenamefont {Ribeiro}\ \emph {et~al.}(2008)\citenamefont
  {Ribeiro}, \citenamefont {Vidal},\ and\ \citenamefont
  {Mosseri}}]{ribeiro2008}%
  \BibitemOpen
  \bibfield  {author} {\bibinfo {author} {\bibfnamefont {P.}~\bibnamefont
  {Ribeiro}}, \bibinfo {author} {\bibfnamefont {J.}~\bibnamefont {Vidal}},\
  and\ \bibinfo {author} {\bibfnamefont {R.}~\bibnamefont {Mosseri}},\ }\href
  {https://doi.org/10.1103/PhysRevE.78.021106} {\bibfield  {journal} {\bibinfo
  {journal} {Phys. Rev. E}\ }\textbf {\bibinfo {volume} {78}},\ \bibinfo
  {pages} {021106} (\bibinfo {year} {2008})}\BibitemShut {NoStop}%
\bibitem [{\citenamefont {Titum}\ and\ \citenamefont
  {Maghrebi}(2020)}]{titum2020}%
  \BibitemOpen
  \bibfield  {author} {\bibinfo {author} {\bibfnamefont {P.}~\bibnamefont
  {Titum}}\ and\ \bibinfo {author} {\bibfnamefont {M.~F.}\ \bibnamefont
  {Maghrebi}},\ }\href {https://doi.org/10.1103/PhysRevLett.125.040602}
  {\bibfield  {journal} {\bibinfo  {journal} {Phys. Rev. Lett.}\ }\textbf
  {\bibinfo {volume} {125}},\ \bibinfo {pages} {040602} (\bibinfo {year}
  {2020})}\BibitemShut {NoStop}%
\bibitem [{\citenamefont {Campa}\ \emph {et~al.}(2009)\citenamefont {Campa},
  \citenamefont {Dauxois},\ and\ \citenamefont
  {Ruffo}}]{campa_statistical_2009}%
  \BibitemOpen
  \bibfield  {author} {\bibinfo {author} {\bibfnamefont {A.}~\bibnamefont
  {Campa}}, \bibinfo {author} {\bibfnamefont {T.}~\bibnamefont {Dauxois}},\
  and\ \bibinfo {author} {\bibfnamefont {S.}~\bibnamefont {Ruffo}},\ }\href
  {https://doi.org/https://doi.org/10.1016/j.physrep.2009.07.001} {\bibfield
  {journal} {\bibinfo  {journal} {Phys. Rep.}\ }\textbf {\bibinfo {volume}
  {480}},\ \bibinfo {pages} {57} (\bibinfo {year} {2009})}\BibitemShut
  {NoStop}%
\bibitem [{\citenamefont {Eisele}\ and\ \citenamefont
  {Ellis}(1988)}]{eisele_multiple_1988}%
  \BibitemOpen
  \bibfield  {author} {\bibinfo {author} {\bibfnamefont {T.}~\bibnamefont
  {Eisele}}\ and\ \bibinfo {author} {\bibfnamefont {R.~S.}\ \bibnamefont
  {Ellis}},\ }\href {https://doi.org/10.1007/BF01016409} {\bibfield  {journal}
  {\bibinfo  {journal} {J. Stat. Phys.}\ }\textbf {\bibinfo {volume} {52}},\
  \bibinfo {pages} {161} (\bibinfo {year} {1988})}\BibitemShut {NoStop}%
\bibitem [{\citenamefont {Canning}(1992)}]{canning_class_1992}%
  \BibitemOpen
  \bibfield  {author} {\bibinfo {author} {\bibfnamefont {A.}~\bibnamefont
  {Canning}},\ }\href
  {https://doi.org/https://doi.org/10.1016/0378-4371(92)90464-2} {\bibfield
  {journal} {\bibinfo  {journal} {Phys. A}\ }\textbf {\bibinfo {volume}
  {185}},\ \bibinfo {pages} {254} (\bibinfo {year} {1992})}\BibitemShut
  {NoStop}%
\bibitem [{\citenamefont {Russomanno}\ \emph {et~al.}(2015)\citenamefont
  {Russomanno}, \citenamefont {Fazio},\ and\ \citenamefont
  {Santoro}}]{russomanno_thermalization_2015}%
  \BibitemOpen
  \bibfield  {author} {\bibinfo {author} {\bibfnamefont {A.}~\bibnamefont
  {Russomanno}}, \bibinfo {author} {\bibfnamefont {R.}~\bibnamefont {Fazio}},\
  and\ \bibinfo {author} {\bibfnamefont {G.~E.}\ \bibnamefont {Santoro}},\
  }\href {https://doi.org/10.1209/0295-5075/110/37005} {\bibfield  {journal}
  {\bibinfo  {journal} {Europhys. Lett.}\ }\textbf {\bibinfo {volume} {110}},\
  \bibinfo {pages} {37005} (\bibinfo {year} {2015})}\BibitemShut {NoStop}%
\bibitem [{\citenamefont {Campbell}\ \emph {et~al.}(2015)\citenamefont
  {Campbell}, \citenamefont {De~Chiara}, \citenamefont {Paternostro},
  \citenamefont {Palma},\ and\ \citenamefont {Fazio}}]{lmg:fidelity}%
  \BibitemOpen
  \bibfield  {author} {\bibinfo {author} {\bibfnamefont {S.}~\bibnamefont
  {Campbell}}, \bibinfo {author} {\bibfnamefont {G.}~\bibnamefont {De~Chiara}},
  \bibinfo {author} {\bibfnamefont {M.}~\bibnamefont {Paternostro}}, \bibinfo
  {author} {\bibfnamefont {G.~M.}\ \bibnamefont {Palma}},\ and\ \bibinfo
  {author} {\bibfnamefont {R.}~\bibnamefont {Fazio}},\ }\href
  {https://doi.org/10.1103/PhysRevLett.114.177206} {\bibfield  {journal}
  {\bibinfo  {journal} {Phys. Rev. Lett.}\ }\textbf {\bibinfo {volume} {114}},\
  \bibinfo {pages} {177206} (\bibinfo {year} {2015})}\BibitemShut {NoStop}%
\bibitem [{\citenamefont {Russomanno}\ \emph {et~al.}(2017)\citenamefont
  {Russomanno}, \citenamefont {Iemini}, \citenamefont {Dalmonte},\ and\
  \citenamefont {Fazio}}]{Russomanno2017}%
  \BibitemOpen
  \bibfield  {author} {\bibinfo {author} {\bibfnamefont {A.}~\bibnamefont
  {Russomanno}}, \bibinfo {author} {\bibfnamefont {F.}~\bibnamefont {Iemini}},
  \bibinfo {author} {\bibfnamefont {M.}~\bibnamefont {Dalmonte}},\ and\
  \bibinfo {author} {\bibfnamefont {R.}~\bibnamefont {Fazio}},\ }\href
  {https://doi.org/10.1103/PhysRevB.95.214307} {\bibfield  {journal} {\bibinfo
  {journal} {Phys. Rev. B}\ }\textbf {\bibinfo {volume} {95}},\ \bibinfo
  {pages} {214307} (\bibinfo {year} {2017})}\BibitemShut {NoStop}%
\bibitem [{\citenamefont {Vu}\ \emph {et~al.}(2022)\citenamefont {Vu},
  \citenamefont {Huang}, \citenamefont {Li},\ and\ \citenamefont
  {Das~Sarma}}]{vu_fermionic_2022}%
  \BibitemOpen
  \bibfield  {author} {\bibinfo {author} {\bibfnamefont {D.~D.}\ \bibnamefont
  {Vu}}, \bibinfo {author} {\bibfnamefont {K.}~\bibnamefont {Huang}}, \bibinfo
  {author} {\bibfnamefont {X.}~\bibnamefont {Li}},\ and\ \bibinfo {author}
  {\bibfnamefont {S.}~\bibnamefont {Das~Sarma}},\ }\href
  {https://doi.org/10.1103/PhysRevLett.128.146601} {\bibfield  {journal}
  {\bibinfo  {journal} {Phys. Rev. Lett.}\ }\textbf {\bibinfo {volume} {128}},\
  \bibinfo {pages} {146601} (\bibinfo {year} {2022})}\BibitemShut {NoStop}%
\bibitem [{\citenamefont {Misguich}\ \emph {et~al.}(2016)\citenamefont
  {Misguich}, \citenamefont {Pasquier},\ and\ \citenamefont
  {Luck}}]{Misguich2016}%
  \BibitemOpen
  \bibfield  {author} {\bibinfo {author} {\bibfnamefont {G.}~\bibnamefont
  {Misguich}}, \bibinfo {author} {\bibfnamefont {V.}~\bibnamefont {Pasquier}},\
  and\ \bibinfo {author} {\bibfnamefont {J.-M.}\ \bibnamefont {Luck}},\ }\href
  {https://doi.org/10.1103/PhysRevB.94.155110} {\bibfield  {journal} {\bibinfo
  {journal} {Phys. Rev. B}\ }\textbf {\bibinfo {volume} {94}},\ \bibinfo
  {pages} {155110} (\bibinfo {year} {2016})}\BibitemShut {NoStop}%
\bibitem [{\citenamefont {Calixto}\ and\ \citenamefont
  {Romera}(2015)}]{calixto_inverse_2015}%
  \BibitemOpen
  \bibfield  {author} {\bibinfo {author} {\bibfnamefont {M.}~\bibnamefont
  {Calixto}}\ and\ \bibinfo {author} {\bibfnamefont {E.}~\bibnamefont
  {Romera}},\ }\href {https://doi.org/10.1088/1742-5468/2015/06/P06029}
  {\bibfield  {journal} {\bibinfo  {journal} {J. Stat. Mech. Theory Exp.}\
  }\textbf {\bibinfo {volume} {2015}},\ \bibinfo {pages} {P06029} (\bibinfo
  {year} {2015})}\BibitemShut {NoStop}%
\bibitem [{\citenamefont {Fujii}(2017)}]{fujii_introduction_2017}%
  \BibitemOpen
  \bibfield  {author} {\bibinfo {author} {\bibfnamefont {K.}~\bibnamefont
  {Fujii}},\ }\href {https://doi.org/10.4236/jmp.2017.812124} {\bibfield
  {journal} {\bibinfo  {journal} {Journal of Modern Physics}\ }\textbf
  {\bibinfo {volume} {8}},\ \bibinfo {pages} {2042} (\bibinfo {year}
  {2017})}\BibitemShut {NoStop}%
\bibitem [{\citenamefont {Sutherland}(2004)}]{Sutherland2004}%
  \BibitemOpen
  \bibfield  {author} {\bibinfo {author} {\bibfnamefont {B.}~\bibnamefont
  {Sutherland}},\ }\href {https://doi.org/10.1142/5552} {\emph {\bibinfo
  {title} {Beautiful Models}}}\ (\bibinfo  {publisher} {WORLD SCIENTIFIC},\
  \bibinfo {year} {2004})\ Chap.~\bibinfo {chapter} {2}\BibitemShut {NoStop}%
\bibitem [{\citenamefont {Abanin}\ \emph {et~al.}(2019)\citenamefont {Abanin},
  \citenamefont {Altman}, \citenamefont {Bloch},\ and\ \citenamefont
  {Serbyn}}]{abanin_colloquium_2019}%
  \BibitemOpen
  \bibfield  {author} {\bibinfo {author} {\bibfnamefont {D.~A.}\ \bibnamefont
  {Abanin}}, \bibinfo {author} {\bibfnamefont {E.}~\bibnamefont {Altman}},
  \bibinfo {author} {\bibfnamefont {I.}~\bibnamefont {Bloch}},\ and\ \bibinfo
  {author} {\bibfnamefont {M.}~\bibnamefont {Serbyn}},\ }\href
  {https://doi.org/10.1103/RevModPhys.91.021001} {\bibfield  {journal}
  {\bibinfo  {journal} {Rev. Mod. Phys.}\ }\textbf {\bibinfo {volume} {91}},\
  \bibinfo {pages} {021001} (\bibinfo {year} {2019})}\BibitemShut {NoStop}%
\bibitem [{\citenamefont {Srednicki}(1994)}]{Srednicki1994}%
  \BibitemOpen
  \bibfield  {author} {\bibinfo {author} {\bibfnamefont {M.}~\bibnamefont
  {Srednicki}},\ }\href {https://doi.org/10.1103/PhysRevE.50.888} {\bibfield
  {journal} {\bibinfo  {journal} {Phys. Rev. E}\ }\textbf {\bibinfo {volume}
  {50}},\ \bibinfo {pages} {888} (\bibinfo {year} {1994})}\BibitemShut
  {NoStop}%
\bibitem [{\citenamefont {Srednicki}(1999)}]{Srednicki_1999}%
  \BibitemOpen
  \bibfield  {author} {\bibinfo {author} {\bibfnamefont {M.}~\bibnamefont
  {Srednicki}},\ }\href {https://doi.org/10.1088/0305-4470/32/7/007} {\bibfield
   {journal} {\bibinfo  {journal} {J. Phys. A: Math. Gen.}\ }\textbf {\bibinfo
  {volume} {32}},\ \bibinfo {pages} {1163} (\bibinfo {year}
  {1999})}\BibitemShut {NoStop}%
\bibitem [{\citenamefont {Holthaus}(2015)}]{holthaus_floquet_2016}%
  \BibitemOpen
  \bibfield  {author} {\bibinfo {author} {\bibfnamefont {M.}~\bibnamefont
  {Holthaus}},\ }\href {https://doi.org/10.1088/0953-4075/49/1/013001}
  {\bibfield  {journal} {\bibinfo  {journal} {J. Phys. B: At. Mol. Opt. Phys.}\
  }\textbf {\bibinfo {volume} {49}},\ \bibinfo {pages} {013001} (\bibinfo
  {year} {2015})}\BibitemShut {NoStop}%
\bibitem [{\citenamefont {Vogl}\ \emph {et~al.}(2020)\citenamefont {Vogl},
  \citenamefont {Rodriguez-Vega},\ and\ \citenamefont
  {Fiete}}]{vogl_effective_2020}%
  \BibitemOpen
  \bibfield  {author} {\bibinfo {author} {\bibfnamefont {M.}~\bibnamefont
  {Vogl}}, \bibinfo {author} {\bibfnamefont {M.}~\bibnamefont
  {Rodriguez-Vega}},\ and\ \bibinfo {author} {\bibfnamefont {G.~A.}\
  \bibnamefont {Fiete}},\ }\href {https://doi.org/10.1103/PhysRevB.101.024303}
  {\bibfield  {journal} {\bibinfo  {journal} {Phys. Rev. B}\ }\textbf {\bibinfo
  {volume} {101}},\ \bibinfo {pages} {024303} (\bibinfo {year}
  {2020})}\BibitemShut {NoStop}%
\bibitem [{\citenamefont {Bukov}\ \emph {et~al.}(2015)\citenamefont {Bukov},
  \citenamefont {D{\textquotesingle}Alessio},\ and\ \citenamefont
  {Polkovnikov}}]{Bukov2014}%
  \BibitemOpen
  \bibfield  {author} {\bibinfo {author} {\bibfnamefont {M.}~\bibnamefont
  {Bukov}}, \bibinfo {author} {\bibfnamefont {L.}~\bibnamefont
  {D{\textquotesingle}Alessio}},\ and\ \bibinfo {author} {\bibfnamefont
  {A.}~\bibnamefont {Polkovnikov}},\ }\href
  {https://doi.org/10.1080/00018732.2015.1055918} {\bibfield  {journal}
  {\bibinfo  {journal} {Adv Phys}\ }\textbf {\bibinfo {volume} {64}},\ \bibinfo
  {pages} {139} (\bibinfo {year} {2015})}\BibitemShut {NoStop}%
\bibitem [{\citenamefont {D'Alessio}\ and\ \citenamefont
  {Rigol}(2014)}]{alessio}%
  \BibitemOpen
  \bibfield  {author} {\bibinfo {author} {\bibfnamefont {L.}~\bibnamefont
  {D'Alessio}}\ and\ \bibinfo {author} {\bibfnamefont {M.}~\bibnamefont
  {Rigol}},\ }\href {https://doi.org/10.1103/PhysRevX.4.041048} {\bibfield
  {journal} {\bibinfo  {journal} {Phys. Rev. X}\ }\textbf {\bibinfo {volume}
  {4}},\ \bibinfo {pages} {041048} (\bibinfo {year} {2014})}\BibitemShut
  {NoStop}%
\bibitem [{\citenamefont {Yousefjani}\ \emph {et~al.}(2023)\citenamefont
  {Yousefjani}, \citenamefont {Bose},\ and\ \citenamefont
  {Bayat}}]{Sougata2023}%
  \BibitemOpen
  \bibfield  {author} {\bibinfo {author} {\bibfnamefont {R.}~\bibnamefont
  {Yousefjani}}, \bibinfo {author} {\bibfnamefont {S.}~\bibnamefont {Bose}},\
  and\ \bibinfo {author} {\bibfnamefont {A.}~\bibnamefont {Bayat}},\ }\href
  {https://doi.org/10.1103/PhysRevResearch.5.013094} {\bibfield  {journal}
  {\bibinfo  {journal} {Phys. Rev. Res.}\ }\textbf {\bibinfo {volume} {5}},\
  \bibinfo {pages} {013094} (\bibinfo {year} {2023})}\BibitemShut {NoStop}%
\bibitem [{\citenamefont {Sierant}\ \emph {et~al.}(2023)\citenamefont
  {Sierant}, \citenamefont {Lewenstein}, \citenamefont {Scardicchio},\ and\
  \citenamefont {Zakrzewski}}]{sierant_2023}%
  \BibitemOpen
  \bibfield  {author} {\bibinfo {author} {\bibfnamefont {P.}~\bibnamefont
  {Sierant}}, \bibinfo {author} {\bibfnamefont {M.}~\bibnamefont {Lewenstein}},
  \bibinfo {author} {\bibfnamefont {A.}~\bibnamefont {Scardicchio}},\ and\
  \bibinfo {author} {\bibfnamefont {J.}~\bibnamefont {Zakrzewski}},\ }\href
  {https://doi.org/10.1103/PhysRevB.107.115132} {\bibfield  {journal} {\bibinfo
   {journal} {Phys. Rev. B}\ }\textbf {\bibinfo {volume} {107}},\ \bibinfo
  {pages} {115132} (\bibinfo {year} {2023})}\BibitemShut {NoStop}%
\bibitem [{\citenamefont {Alet}\ and\ \citenamefont
  {Laflorencie}(2018)}]{Fabien2018}%
  \BibitemOpen
  \bibfield  {author} {\bibinfo {author} {\bibfnamefont {F.}~\bibnamefont
  {Alet}}\ and\ \bibinfo {author} {\bibfnamefont {N.}~\bibnamefont
  {Laflorencie}},\ }\href
  {https://doi.org/https://doi.org/10.1016/j.crhy.2018.03.003} {\bibfield
  {journal} {\bibinfo  {journal} {C. R. Phys.}\ }\textbf {\bibinfo {volume}
  {19}},\ \bibinfo {pages} {498} (\bibinfo {year} {2018})}\BibitemShut
  {NoStop}%
\bibitem [{\citenamefont {Garratt}\ and\ \citenamefont
  {Roy}(2022)}]{garratt_resonant_2022}%
  \BibitemOpen
  \bibfield  {author} {\bibinfo {author} {\bibfnamefont {S.~J.}\ \bibnamefont
  {Garratt}}\ and\ \bibinfo {author} {\bibfnamefont {S.}~\bibnamefont {Roy}},\
  }\href {https://doi.org/10.1103/PhysRevB.106.054309} {\bibfield  {journal}
  {\bibinfo  {journal} {Phys. Rev. B}\ }\textbf {\bibinfo {volume} {106}},\
  \bibinfo {pages} {054309} (\bibinfo {year} {2022})}\BibitemShut {NoStop}%
\bibitem [{\citenamefont {Stinchcombe}(1973)}]{stinchcombe_ising_1973}%
  \BibitemOpen
  \bibfield  {author} {\bibinfo {author} {\bibfnamefont {R.~B.}\ \bibnamefont
  {Stinchcombe}},\ }\href {https://doi.org/10.1088/0022-3719/6/15/009}
  {\bibfield  {journal} {\bibinfo  {journal} {J. Phys. C: Solid State Phys.}\
  }\textbf {\bibinfo {volume} {6}},\ \bibinfo {pages} {2459} (\bibinfo {year}
  {1973})}\BibitemShut {NoStop}%
\bibitem [{\citenamefont {George~Arfken}(2011)}]{arfkenmath}%
  \BibitemOpen
  \bibfield  {author} {\bibinfo {author} {\bibfnamefont {F.~E.~H.}\
  \bibnamefont {George~Arfken}, \bibfnamefont {Hans~Weber}},\ }\href
  {https://www.elsevier.com/books/mathematical-methods-for-
  physicists/arfken/978-0-12-384654-9} {\emph {\bibinfo {title} {Mathematical
  Methods for Physicists}}},\ \bibinfo {edition} {7th}\ ed.\ (\bibinfo
  {publisher} {Academic Press},\ \bibinfo {year} {2011})\BibitemShut {NoStop}%
\bibitem [{\citenamefont {Johansson}\ \emph {et~al.}(2013)\citenamefont
  {Johansson}, \citenamefont {Nation},\ and\ \citenamefont {Nori}}]{qutip}%
  \BibitemOpen
  \bibfield  {author} {\bibinfo {author} {\bibfnamefont {J.}~\bibnamefont
  {Johansson}}, \bibinfo {author} {\bibfnamefont {P.}~\bibnamefont {Nation}},\
  and\ \bibinfo {author} {\bibfnamefont {F.}~\bibnamefont {Nori}},\ }\href
  {https://doi.org/https://doi.org/10.1016/j.cpc.2012.11.019} {\bibfield
  {journal} {\bibinfo  {journal} {Comput. Phys. Commun.}\ }\textbf {\bibinfo
  {volume} {184}},\ \bibinfo {pages} {1234} (\bibinfo {year}
  {2013})}\BibitemShut {NoStop}%
\bibitem [{\citenamefont {Rahaman}\ \emph {et~al.}(2024)\citenamefont
  {Rahaman}, \citenamefont {Sakurai},\ and\ \citenamefont
  {Roy}}]{rahaman2024time}%
  \BibitemOpen
  \bibfield  {author} {\bibinfo {author} {\bibfnamefont {M.}~\bibnamefont
  {Rahaman}}, \bibinfo {author} {\bibfnamefont {A.}~\bibnamefont {Sakurai}},\
  and\ \bibinfo {author} {\bibfnamefont {A.}~\bibnamefont {Roy}},\ }\href@noop
  {} {\bibinfo {title} {Time crystal embodies chimera in periodically driven
  quantum spin system}} (\bibinfo {year} {2024}),\ \Eprint
  {https://arxiv.org/abs/2309.16523} {arXiv:2309.16523 [cond-mat.stat-mech]}
  \BibitemShut {NoStop}%
\bibitem [{\citenamefont {Mukherjee}\ \emph {et~al.}(2015)\citenamefont
  {Mukherjee}, \citenamefont {Spracklen}, \citenamefont {Choudhury},
  \citenamefont {Goldman}, \citenamefont {Öhberg}, \citenamefont {Andersson},\
  and\ \citenamefont {Thomson}}]{mukherjee_modulation-assisted_2015}%
  \BibitemOpen
  \bibfield  {author} {\bibinfo {author} {\bibfnamefont {S.}~\bibnamefont
  {Mukherjee}}, \bibinfo {author} {\bibfnamefont {A.}~\bibnamefont
  {Spracklen}}, \bibinfo {author} {\bibfnamefont {D.}~\bibnamefont
  {Choudhury}}, \bibinfo {author} {\bibfnamefont {N.}~\bibnamefont {Goldman}},
  \bibinfo {author} {\bibfnamefont {P.}~\bibnamefont {Öhberg}}, \bibinfo
  {author} {\bibfnamefont {E.}~\bibnamefont {Andersson}},\ and\ \bibinfo
  {author} {\bibfnamefont {R.~R.}\ \bibnamefont {Thomson}},\ }\href
  {https://doi.org/10.1088/1367-2630/17/11/115002} {\bibfield  {journal}
  {\bibinfo  {journal} {New J. Phys.}\ }\textbf {\bibinfo {volume} {17}},\
  \bibinfo {pages} {115002} (\bibinfo {year} {2015})}\BibitemShut {NoStop}%
\bibitem [{\citenamefont {Lin}\ \emph {et~al.}(2018)\citenamefont {Lin},
  \citenamefont {Sbierski}, \citenamefont {Dorfner}, \citenamefont {Karrasch},\
  and\ \citenamefont {Heidrich-Meisner}}]{lin_many-body_2018}%
  \BibitemOpen
  \bibfield  {author} {\bibinfo {author} {\bibfnamefont {S.-H.}\ \bibnamefont
  {Lin}}, \bibinfo {author} {\bibfnamefont {B.}~\bibnamefont {Sbierski}},
  \bibinfo {author} {\bibfnamefont {F.}~\bibnamefont {Dorfner}}, \bibinfo
  {author} {\bibfnamefont {C.}~\bibnamefont {Karrasch}},\ and\ \bibinfo
  {author} {\bibfnamefont {F.}~\bibnamefont {Heidrich-Meisner}},\ }\href
  {https://doi.org/10.21468/SciPostPhys.4.1.002} {\bibfield  {journal}
  {\bibinfo  {journal} {SciPost Phys.}\ }\textbf {\bibinfo {volume} {4}},\
  \bibinfo {pages} {002} (\bibinfo {year} {2018})}\BibitemShut {NoStop}%
\bibitem [{\citenamefont {Murphy}\ \emph {et~al.}(2011)\citenamefont {Murphy},
  \citenamefont {Wortis},\ and\ \citenamefont
  {Atkinson}}]{murphy_generalized_2011}%
  \BibitemOpen
  \bibfield  {author} {\bibinfo {author} {\bibfnamefont {N.~C.}\ \bibnamefont
  {Murphy}}, \bibinfo {author} {\bibfnamefont {R.}~\bibnamefont {Wortis}},\
  and\ \bibinfo {author} {\bibfnamefont {W.~A.}\ \bibnamefont {Atkinson}},\
  }\href {https://doi.org/10.1103/PhysRevB.83.184206} {\bibfield  {journal}
  {\bibinfo  {journal} {Phys. Rev. B}\ }\textbf {\bibinfo {volume} {83}},\
  \bibinfo {pages} {184206} (\bibinfo {year} {2011})}\BibitemShut {NoStop}%
\bibitem [{\citenamefont {Torres-Herrera}\ \emph {et~al.}(2020)\citenamefont
  {Torres-Herrera}, \citenamefont {Vallejo-Fabila}, \citenamefont
  {Mart\'{\i}nez-Mendoza},\ and\ \citenamefont
  {Santos}}]{torres-herrera_self-averaging_2020}%
  \BibitemOpen
  \bibfield  {author} {\bibinfo {author} {\bibfnamefont {E.~J.}\ \bibnamefont
  {Torres-Herrera}}, \bibinfo {author} {\bibfnamefont {I.}~\bibnamefont
  {Vallejo-Fabila}}, \bibinfo {author} {\bibfnamefont {A.~J.}\ \bibnamefont
  {Mart\'{\i}nez-Mendoza}},\ and\ \bibinfo {author} {\bibfnamefont {L.~F.}\
  \bibnamefont {Santos}},\ }\href {https://doi.org/10.1103/PhysRevE.102.062126}
  {\bibfield  {journal} {\bibinfo  {journal} {Phys. Rev. E}\ }\textbf {\bibinfo
  {volume} {102}},\ \bibinfo {pages} {062126} (\bibinfo {year}
  {2020})}\BibitemShut {NoStop}%
\bibitem [{\citenamefont {Trivedi}\ and\ \citenamefont
  {Heidarian}(2005)}]{trivedi_can_2005}%
  \BibitemOpen
  \bibfield  {author} {\bibinfo {author} {\bibfnamefont {N.}~\bibnamefont
  {Trivedi}}\ and\ \bibinfo {author} {\bibfnamefont {D.}~\bibnamefont
  {Heidarian}},\ }\href {https://doi.org/10.1143/PTPS.160.296} {\bibfield
  {journal} {\bibinfo  {journal} {Prog Theor Phys Supp}\ }\textbf {\bibinfo
  {volume} {160}},\ \bibinfo {pages} {296} (\bibinfo {year}
  {2005})}\BibitemShut {NoStop}%
\bibitem [{\citenamefont {Defenu}\ \emph {et~al.}(2018)\citenamefont {Defenu},
  \citenamefont {Enss}, \citenamefont {Kastner},\ and\ \citenamefont
  {Morigi}}]{defenu2018}%
  \BibitemOpen
  \bibfield  {author} {\bibinfo {author} {\bibfnamefont {N.}~\bibnamefont
  {Defenu}}, \bibinfo {author} {\bibfnamefont {T.}~\bibnamefont {Enss}},
  \bibinfo {author} {\bibfnamefont {M.}~\bibnamefont {Kastner}},\ and\ \bibinfo
  {author} {\bibfnamefont {G.}~\bibnamefont {Morigi}},\ }\href
  {https://doi.org/10.1103/PhysRevLett.121.240403} {\bibfield  {journal}
  {\bibinfo  {journal} {Phys. Rev. Lett.}\ }\textbf {\bibinfo {volume} {121}},\
  \bibinfo {pages} {240403} (\bibinfo {year} {2018})}\BibitemShut {NoStop}%
\bibitem [{\citenamefont {Mori}(2019)}]{mori_prethermalization_2019}%
  \BibitemOpen
  \bibfield  {author} {\bibinfo {author} {\bibfnamefont {T.}~\bibnamefont
  {Mori}},\ }\href {https://doi.org/10.1088/1751-8121/aaf9db} {\bibfield
  {journal} {\bibinfo  {journal} {J. Phys. A: Math. Theor.}\ }\textbf {\bibinfo
  {volume} {52}},\ \bibinfo {pages} {054001} (\bibinfo {year}
  {2019})}\BibitemShut {NoStop}%
\bibitem [{\citenamefont {Sciolla}\ and\ \citenamefont
  {Biroli}(2010)}]{sciolla_quantum_2010}%
  \BibitemOpen
  \bibfield  {author} {\bibinfo {author} {\bibfnamefont {B.}~\bibnamefont
  {Sciolla}}\ and\ \bibinfo {author} {\bibfnamefont {G.}~\bibnamefont
  {Biroli}},\ }\href {https://doi.org/10.1103/PhysRevLett.105.220401}
  {\bibfield  {journal} {\bibinfo  {journal} {Phys. Rev. Lett.}\ }\textbf
  {\bibinfo {volume} {105}},\ \bibinfo {pages} {220401} (\bibinfo {year}
  {2010})}\BibitemShut {NoStop}%
\bibitem [{\citenamefont {Kidd}\ \emph {et~al.}(2019)\citenamefont {Kidd},
  \citenamefont {Olsen},\ and\ \citenamefont {Corney}}]{Kidd2019}%
  \BibitemOpen
  \bibfield  {author} {\bibinfo {author} {\bibfnamefont {R.~A.}\ \bibnamefont
  {Kidd}}, \bibinfo {author} {\bibfnamefont {M.~K.}\ \bibnamefont {Olsen}},\
  and\ \bibinfo {author} {\bibfnamefont {J.~F.}\ \bibnamefont {Corney}},\
  }\href {https://doi.org/10.1103/PhysRevA.100.013625} {\bibfield  {journal}
  {\bibinfo  {journal} {Phys. Rev. A}\ }\textbf {\bibinfo {volume} {100}},\
  \bibinfo {pages} {013625} (\bibinfo {year} {2019})}\BibitemShut {NoStop}%
\bibitem [{\citenamefont {B\"acker}\ \emph {et~al.}(2004)\citenamefont
  {B\"acker}, \citenamefont {F\"urstberger},\ and\ \citenamefont
  {Schubert}}]{husimi}%
  \BibitemOpen
  \bibfield  {author} {\bibinfo {author} {\bibfnamefont {A.}~\bibnamefont
  {B\"acker}}, \bibinfo {author} {\bibfnamefont {S.}~\bibnamefont
  {F\"urstberger}},\ and\ \bibinfo {author} {\bibfnamefont {R.}~\bibnamefont
  {Schubert}},\ }\href {https://doi.org/10.1103/PhysRevE.70.036204} {\bibfield
  {journal} {\bibinfo  {journal} {Phys. Rev. E}\ }\textbf {\bibinfo {volume}
  {70}},\ \bibinfo {pages} {036204} (\bibinfo {year} {2004})}\BibitemShut
  {NoStop}%
\bibitem [{\citenamefont {Sachdev}(2011)}]{sachdev_quantum_2011}%
  \BibitemOpen
  \bibfield  {author} {\bibinfo {author} {\bibfnamefont {S.}~\bibnamefont
  {Sachdev}},\ }\href {https://doi.org/10.1017/CBO9780511973765} {\emph
  {\bibinfo {title} {Quantum {Phase} {Transitions}}}},\ \bibinfo {edition}
  {2nd}\ ed.\ (\bibinfo  {publisher} {Cambridge University Press},\ \bibinfo
  {year} {2011})\BibitemShut {NoStop}%
\bibitem [{Note1()}]{Note1}%
  \BibitemOpen
  \bibinfo {note} {Thermodynamically, at $\beta \equiv 1/k_BT=0$, we have
  $\expval {H}^2_{std}=\Tr {H^2_0}/2^N$. Substituting eqn.~\ref {eq:h0h1} for
  $H_0$ yields $1/8$ on the RHS.}\BibitemShut {Stop}%
\bibitem [{Note2()}]{Note2}%
  \BibitemOpen
  \bibinfo {note} {When frozen, $\ket {\psi (t)}\approx \ket {s_N}$. From
  Eq.~(\ref {eq:driven:ham}), $\expval {\protect \hat {H}_{0,1}}$ are both
  approx. constant. Averaging the square of $\expval {\protect \hat {H}(t)}$
  over long times yields dependency $\sim h^2 + \delta $, where $\delta \sim
  h_0\ll 1$. Thus, the std. devn. in time $\sim h\sim \omega $, since $\eta
  =4h/\omega $ is kept fixed.}\BibitemShut {Stop}%
\end{thebibliography}%

\end{document}